\documentclass[10pt,journal,compsoc]{IEEEtran}

\ifCLASSOPTIONcompsoc
\usepackage[nocompress]{cite}
\else
\usepackage{cite}
\fi

\ifCLASSINFOpdf

\else

\fi

\usepackage{graphicx}
\usepackage{xcolor}
\usepackage{array}
\usepackage{caption}
\begin{document}
	
	\title{The Dichotomy of Cloud and IoT:  Cloud-Assisted IoT From a Security Perspective}

	\author{Behrouz Zolfaghari, Abbas Yazdinejad, Ali Dehghantanha, Jacob Krzciok, and Khodakhast Bibak
		
		\thanks{B.~Zolfaghari, A.~Yazdinejad, and A.~Dehghantanha are with the Cyber Science Lab, School of Computer Science, University of Guelph, Guelph, Ontario, Canada. Email: {\tt behrouz@cybersciencelab.org, ayazdine@uoguelph.ca, adehghan@uoguelph.ca}. J.~Krzciok and K.~Bibak are with the Department of Computer Science and Software Engineering, Miami University, Oxford, Ohio, 45056, USA. Email: {\tt \{krzciojj,bibakk\}@miamioh.edu}}}

	

	
	\IEEEtitleabstractindextext{%
		\begin{abstract}
			\textcolor{black}{In recent years, the existence of a significant cross-impact between Cloud computing and Internet of Things (IoT) has lead to a dichotomy that gives raise to \textit{Cloud-Assisted IoT (CAIoT)} and \textit{IoT-Based Cloud (IoTBC)}. Although it is pertinent to study both technologies, this paper focuses on CAIoT, and especially its security issues, which are inherited from both Cloud computing and IoT. This study starts with reviewing existing relevant surveys, noting their shortcomings, which motivate a comprehensive survey in this area. We proceed to highlight existing approaches towards the design of Secure CAIoT (SCAIoT) along with related security challenges and controls. We develop a layered architecture for SCAIoT. Furthermore, we take a look at what the future may hold for SCAIoT with a focus on the role of Artificial Intelligence(AI).}
			
		\end{abstract}
		
		}

	\maketitle

	\IEEEdisplaynontitleabstractindextext
	
	\IEEEpeerreviewmaketitle

	\section{Introduction}\label{sec:introduction}

	\textcolor{black}{In recent years, cloud computing has been of great interest to researchers \cite{Into-Jour001}\cite{Into-Jour002}. Its role in providing on-demand services and resources has opened its way into a variety of technological environments, ranging from data centers~\cite{New-New-Cloud-Jour001}, power systems~\cite{New-New-Cloud-Jour001} and video delivery systems~\cite{New-New-Cloud-Jour001} to earthquake command systems~\cite{Into-Jour003} and intelligent transportation~\cite{New-New-Cloud-Jour001}\cite{a2}.}

	
	A variety of design objectives including fairness~\cite{Into-Jour007}, fault tolerance~\cite{Into-Jour008}, energy consumption~\cite{Into-Jour009,a3} and reliability~\cite{Into-Jour010} are considered in the design of cloud computing systems. \textcolor{black}{However, security is probably the most critical design objective in this field \cite{Into-Jour013}\cite{Into-Jour012}\cite{Into-Jour015}\cite{Into-Jour014,Into-Jour016}\cite{a1}.} 
	

	
	\textcolor{black}{IoT frequently appears in the ecosystem of Cloud computing. This technology integrates geographically-distributed cyber-physical devices or cyber-enabled systems with the goal of providing strategic services\cite{a4, a5}.} The application areas of IoT vary from transport and healthcare to agriculture and FinTech. \textcolor{black}{Similar to the case of Cloud computing, security is the most critical objective in the design of IoT systems \cite{Into-Jour025}\cite{Into-Jour026}\cite{abc}\cite{Into-Jour028}.}
	
\begin{figure} 
		\centering
		\includegraphics[width=1\linewidth,keepaspectratio]{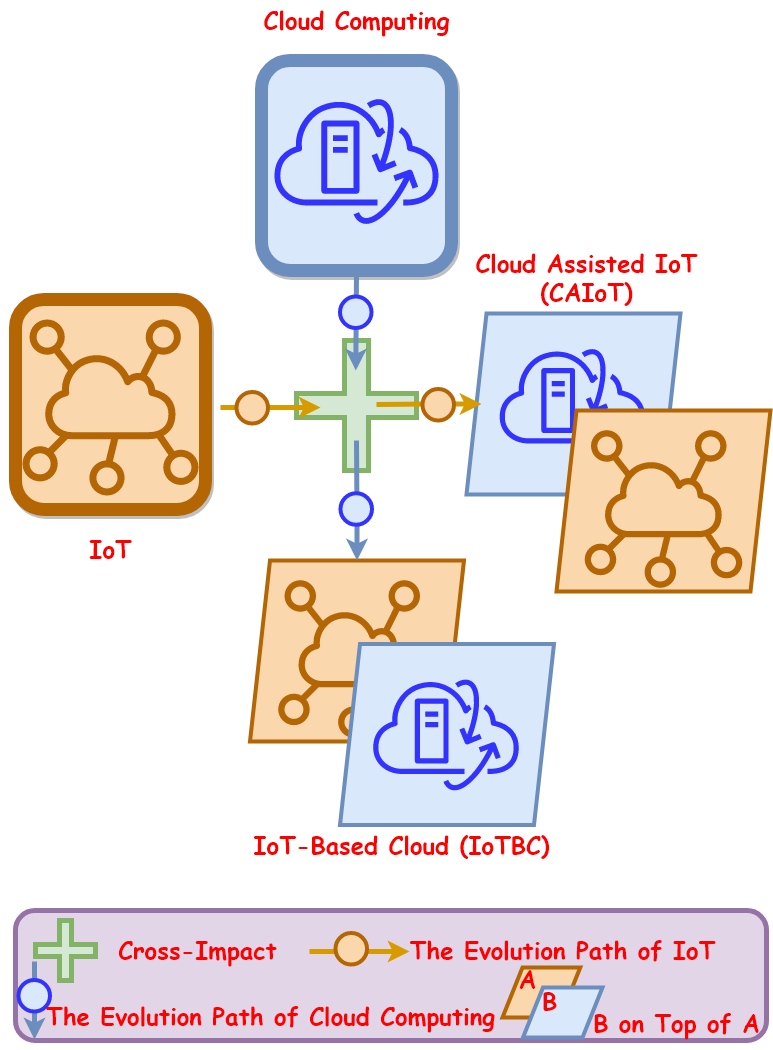}
		\caption{\textcolor{black}{The Dichotomy of Cloud Computing and IoT}}
		\label{fig:Cloud Assisted IoT}
	\end{figure}

	
	\textcolor{black}{Recent literature reveals a cross-impact between IoT and Cloud computing, which leads to an inseparable dichotomy. This dichotomy is illustrated in Figure~\ref{fig:Cloud Assisted IoT}.} 
	
	\textcolor{black}{Figure \ref{fig:Cloud Assisted IoT} introduces the icons we will use in the rest of this paper for Cloud, IoT, CAIoT and IoTBC. In this figure, overlapping parallelograms represent technologies on top of each other. As seen in the figure, the dichotomy of Cloud and IoT leads to two emerging technologies, namely CAIoT and IoTBC. CAIoT deploys IoT on top of Cloud, while in IoTBC, Cloud services are provided leveraging IoT capabilities.}
	
	
	
\begin{figure} 
		\centering
		\includegraphics[width=0.99\linewidth,keepaspectratio]{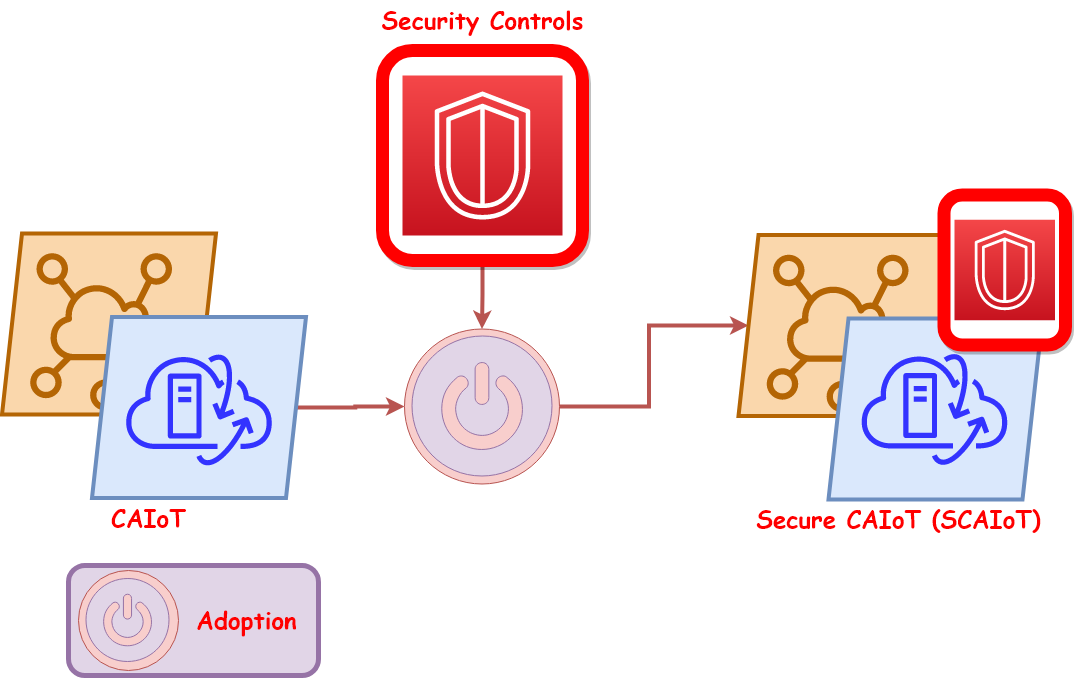}
		\caption{\textcolor{black}{SCAIoT: CAIoT with Adopted Security Controls}}
		\label{fig:Secure Cloud Assisted IoT}
	\end{figure}

	\textcolor{black}{The dichotomy of Figure \ref{fig:Cloud Assisted IoT} needs to be studied form both sides; CAIoT and IoTBC, each of which brings about a variety of challenges, issues and considerations. A comprehensive survey on each of the technologies can pave the way for further research. In this survey, we study CAIoT from a security point of view. The adoption of security controls by CAIoT gives raise to SCAIoT. The notion of SCAIoT is demonstrated in Figure \ref{fig:Secure Cloud Assisted IoT}. (introduced icons will be used in the rest of this study to represent SCAIoT)} 
	
	\textcolor{black}{We present a survey on the literature of SCAIoT. We identify existing approaches towards the design of SCAIoT. We highlight security challenges faced by each approach. Moreover, we study the security controls used to address the challenges. We establish a layered architecture for SCAIoT, which reflects all the identified approaches. Furthermore, we develop a future roadmap for SCAIoT with a focus on the role of AI in the present and the future of this technology.} 
	
	\textcolor{black}{Although there are some relevant surveys, there are some shortcoming within them (Discussed in Subsection \ref{Motiv}), which motivate our work in this paper.}

	\textcolor{black}{In this review, we seek to answer the following questions:}
	
	\begin{itemize}
		\item \textcolor{black}{What are the security challenges of SCAIoT? A security challenge is a requirement needed to be met for a system to be secure.}
		
		\item \textcolor{black}{What are the the security controls need to be accomplished in SCAIoT? A security control is a procedure or a program run in order to deploy, improve or sustain security in a system.}
		
		\item \textcolor{black}{What should the architecture of SCAIoT look like?}

		\item \textcolor{black}{What roles does AI play in the future of SCAIoT?}
		
	\end{itemize}
	
	
	\textcolor{black}{In answering the above questions, we offer the following contributions: }

	\begin{itemize}
				
		\item \textcolor{black}{We identify the following four approaches towards the design of  SCAIoT (Discussed in Section \ref{AppChall}}).
		
		\begin{itemize}
			\item \textbf{\textcolor{black}{IoT on Top of Secure Cloud (Subsection \ref{Appr-Sub1}):}}
			In this approach, system designers handle security issues in the cloud layer, while taking IoT requirements into consideration.

			\item \textbf{\textcolor{black}{Secure IoT on Top of Cloud (Subsection \ref{Appr-Sub2}):}}
			In this approach, security requirements are provided by the IoT system with consideration of Cloud requirements.

			\item \textbf{Secure IoT-Cloud Communications (Subsection \ref{Appr-Sub3}):}
			The system designer attempts to protect the communication between the two layers from security threats.
			
			\item \textbf{Integrated \textcolor{black}{SCAIoT} Solutions (Subsection \ref{Appr-Sub4}):}
			Security provisions are applied to the CAIoT as a whole \textcolor{black}{via secure codesign, interfacing and integration}.
		\end{itemize}
		
		\item \textcolor{black}{We identify the challenges faced by each of the aforementioned approaches (Summarized in Subsection \ref{Summ-Chall}).} 
		
		\item \textcolor{black}{We identify the security controls used by each of these approaches (Summarized in Subsection \ref{Summ-Cont}).}
		
		\item \textcolor{black}{We establish a layered architecture for SCAIoT (subsection \ref{Lay-Arch}.)}
		
		\item \textcolor{black}{We attempt to anticipate the impact of AI and related advancements on the future of research in this field. We contribute a future roadmap for further research in this area (Section \ref{Fut}).}
		
	\end{itemize}
	
	\textcolor{black}{The rest of this paper is organized as follows. Section~\ref{ExSurv} reviews  existing surveys and sheds light on their shortcomings to highlight our motivations for the work of this paper. Section~\ref{AppChall} discusses approaches towards CAIoT security, along with the related challenges and controls. Section \ref{Disc} summarizes the identified challenges and controls, and establishes the layered architecture. Section~\ref{Fut} presents the future roadmap. Section~\ref{Conc} concludes the paper.}
	
	\section{Existing Surveys}\label{ExSurv}
	
	Relevant existing surveys are briefly reviewed in this section by publication year and are categorized based on the role of AI, future roadmap, security coverage, and strategies they focused on.

	subsection{Surveys on Cloud-Assisted IoT}
	Several research surveys has been conducted on cloud-assisted IoT. 
	The benefit of these surveys is that they enable the researchers to understand the challenges to overcome to implement secure CAIoT networks and applications. 
	
	The authors of~\cite{Surve-Conf006} presented a vision to merge cloud computing and IoT. An infrastructure based on IoT Cloud was designed to mitigate the challenges associated with the integration of IoT and Cloud Computing. A similar system to overcome the integration obstacles and to improve the Quality of Service (QoS) has been proposed in~\cite{Surve-Conf007}. To deal with the issue of efficient data storage and retrieval, a similar framework  was  developed in~\cite{Surve-Conf003} where fog computing and cloud computing were implemented. Moreover,~\cite{Surve-Conf003} surveys the benefits and shortcomings of various IoT-based cloud models. Resource management was further studied in~\cite{Surve-Conf010} where the challenges involved in the modeling and deployment of IoT applications and their mitigation were discussed.
	
	The authors of~\cite{Surve-Conf001} discuss the rising complexity of IoT frameworks used in applications, enterprises, and business. They argue that traditional architectures have to be modified to handle upcoming challenges. Thus, a comprehensive study on IoT framework architectures and their performance is presented. In~\cite{Surve-Conf005}, multiple architectures integrating FOG computing, EDGE computing, and mobile cloud computing have been reviewed and a taxonomy has been proposed. 
	For instance, \cite{Surve-Conf011} studies cloud-based IoT-enabled smart grids.
	
	Girs, at el.~\cite{Surv-Jour001} performed a systematic literature review on Service Level Agreement~(SLAs) related to cloud computing and the IoT ecosystem. This work studies the management of SLAs for cloud computing and IoT. It identifies relevant literature and flaws and discusses opportunities for cloud services in IoT.  In~\cite{Surve-Conf005}, the authors present a taxonomy to survey the existing IoT-cloud architectures and classifies based on various criteria.
	
	In relevant work, Sikarwar, et al.~\cite{Surve-Conf003} review IoT-enabled cloud platforms used in the integration with IoT. In addition, they examine the various applications of fog computing. The seamless data exchange between IoT devices and sensors requires outlets for controlling, storing, and analyzing data \cite{c5}. Thus, cloud computing and fog computing were used interchangeably to provide high storage capacity and processing capability. 

	Cloud-based IoT has a disadvantage of major power consumption. Hasan, el at.~\cite{Surve-Conf008} review research on energy-efficient schemes for IoT in various cloud environments. They investigate the results according to previous methods, and they discuss their shortcomings and possible opportunities which can be investigated in the future to improve these models.
	
	In~\cite{Surve-Conf009}, the authors primarily discuss the procedure behind handling sensor observation streams in cloud-based IoT platforms. 
	They present a method that delivers a complete \textcolor{black}{solution} that considers IoT platforms'  challenges and non-functional requirements, such as platform adaptation, scalability, and availability. In a related survey, Tata, at el.~\cite{Surve-Conf010} investigated opportunities and challenges of IoT application architecture. Research and development challenges in the lifecycle of IoT applications, including modeling, deployment, and support of non-functional requirements such as security, privacy, performance and provenance, were investigated. In this work, the authors summarize the challenges related to the modeling and deployment of IoT applications and probable research directions in overcoming these challenges. The authors of~\cite{Surve-Conf004} research Platform as a service (PaaS) aspects of IoT end-user applications. They have critically reviewed and discussed PaaS on the whole spectrum of IoT verticals.

	\subsection{Surveys on Secure Cloud-Assisted IoT}
	Achieving security in cloud-assisted IoT environments has been a challenging task \cite{a6,a7}. 
	\textcolor{black}{Due} to this, various surveys related to security challenges such as intrusion detection, threat modeling, etc. were published. Public-key encryption with search functionality is one of the widely used cryptographic techniques that help in data retrieval without decryption. It consists of public-key encryption with keyword search (PKE-ES), public-key encryption with equality test (PKE-ET), and plaintext checkable encryption (PCE). The authors of \cite{Surv-Jour002} examine PKE-SF design in terms of security and performance. Moreover, an architecture for next-generation mobile technologies on Cloud-based IoT that provide privacy-preserving data aggregation without public-key encryption is presented in~\cite{Surv-Jour004}. This paper presents several open problems with promising ideas to stimulate more research on Cloud-Based IoT Security and Privacy.
	
	The author of~\cite{Surve-Conf013} study IoT data security. This work Combines the concept of Cloud-based IoT architecture with different layers toward proposes different security issues and challenges in IoT. Alsaidi, et al.~\cite{Surve-Conf012} proposed an architecture of Cloud-assisted IoT applications for smart cities, telemedicine, and intelligent transportation systems. By and large, authors investigate the security threats and attacks introduced by unauthorized access and misuse of information collected by IoT devices.

	Cyber-Physical Systems (CPS) are regarded as  intelligent industrial system, is widely used in intelligent transportation \cite{a9}, smart grids, eHealthcare \cite{a8}, etc. These systems primarily use Supervisory Control and Data Acquisition~(SCADA) to monitor and control their critical infrastructure. However, like IoT, these systems pose significant security and privacy challenges. The authors of~\cite{Surv-Jour003} propose a study related to security challenges that an industrial system faces in an IoT cloud supported environment and  highlight the security challenges prevailing in classical SCADA systems. 
	
	\subsection{\textcolor{black}{Summary}}
	
	\textcolor{black}{Table~\ref{table:Existing Surveys} summarizes the  surveys reviewed in this section in order to make it easier to compare them with our work in this paper.}
	
	\captionsetup[table]{justification=justified,singlelinecheck=off,  labelfont=bf}
	
	\begin{table}
		\centering
		\caption{\textcolor{black}{Summary of Existing Surveys}}
		\label{table:Existing Surveys}
		\begin{tabular}{|c|c|c|c|c|c|c|}
			\hline
			\textbf{Survey} & \textbf{Year} &\textbf{CAIoT} &\textbf{Security}  &\textbf{Lay. Arch.} &\textbf{Roadmap} & \textbf{AI} \\ \hline
			{\cite{Surve-Conf006}}     & 2021   & No      & No                & No         & No               & No            \\ \hline
			{\cite{Surve-Conf007}}     & 2021    & Yes      & Yes              & No            & No              & No           \\ \hline
			{\cite{Surve-Conf001}}     & 2021    & No     & No                  & Yes        & No              & No           \\ \hline
			{ \cite{Surv-Jour001}}     & 2020    & No     & No                   & No       & No              & No          \\ \hline
			{\cite{Surve-Conf005}}     & 2020    & Yes      & No                 & Yes         & No              & No          \\ \hline
			{\cite{Surve-Conf003}}     & 2020          & No                     & No &  No      & Yes              & No         \\ \hline
			{\cite{Surve-Conf011}}     & 2020     & Yes     & Yes                &No          & No              & No         \\ \hline
			{\cite{Surve-Conf009}}     & 2017   & Yes       & Yes               & No           & No              & No         \\ \hline
			{\cite{Surve-Conf010}}     & 2017    & Yes      & Yes                 & Yes        & Yes               & No        \\ \hline
			{\cite{Surve-Conf004}}     & 2016     & Yes     & No                  & No        & No            & No           \\ \hline
			{\cite{Surv-Jour002}}     & 2021     & No     & Yes                   & No      & No               & No         \\ \hline
			{ \cite{Surve-Conf013}}     & 2021    & No      & Yes                  & No       & No               & No      \\ \hline
			{\cite{Surve-Conf012}}     & 2018    & No      & Yes                 & No        & No              & No        \\ \hline
			{\cite{Surv-Jour004}}     & 2017     & No     & Yes                   & No      & Yes              & No          \\ \hline
			{\cite{Surv-Jour003}}     & 2016     & No     & Yes                    & No     & Yes              & No          \\ \hline
			{ \cite{Surve-Conf018}}     & 2020     & Yes     & Yes                  & No       & Yes              & No         \\ \hline
			{\cite{Surve-Conf017}}     & 2020      & Yes    & Yes                   & No      & No              & No           \\ \hline
			{\cite{Surve-Conf016}}     & 2019    & Yes      & Yes                 & No        & No              & No           \\ \hline
			{\cite{Surve-Conf019}}     & 2019     & No     & Yes                & No         & No              & No           \\ \hline
			{\cite{Surve-Conf014}}     & 2018     & Yes     & Yes                  & No       & No              & Yes          \\ \hline
			{\cite{Surve-Conf015}}     & 2017     & Yes     & Yes                 & No        & No              & No           \\ \hline
			{\cite{Surve-Conf017-1}}     & 2016    & Yes     & Yes                & No         & No             & No           \\ \hline
		\end{tabular}
	\end{table}
	
	
	
	\textcolor{black}{In Table \ref{table:Existing Surveys}, the first entry in each row cites one of the reviewed surveys. The second column contains the publication year. This column helps the reader identify outdated surveys.The third column contains "Yes" if the related survey covers CAIoT. It contains "No" otherwise. The fifth column indicates whether the survey discusses security or not. The sixth column contains a "Yes" for surveys that present a future roadmap. The seventh column shows whether the survey studies the role of AI in the future roadmap or not.}

	\subsection{Motivations}\label{Motiv}
	
	\textcolor{black}{As seen Table \ref{table:Existing Surveys}, although there are a number of surveys somehow related to CAIoT and related areas, There is no survey with all of the following features.}
	
	\begin{itemize}
		\item \textcolor{black}{Published recently}
		
		\item \textcolor{black}{Focusing on CAIoT}
		
		\item \textcolor{black}{Focusing on security}
		
		\item \textcolor{black}{Developing a layered architecture}
		
		\item \textcolor{black}{Presenting a future roadmap}
		
		\item \textcolor{black}{Studying the role of AI}
	\end{itemize}
	
	\textcolor{black}{The main motivation of this review is to address the above gap.}
	
	\section{Approaches and Challenges}\label{AppChall}
	This section is dedicated to the approaches and challenges prevailing in IoT over cloud and CAIoT. Most security and privacy attacks target data during transmission between IoT devices and the cloud. To address this, various approaches and their associated challenges are discussed in Subsections 3.1 - 3.4.
	
	\subsection{\textcolor{black}{IoT on Top of Secure Cloud}}\label{Appr-Sub1}
\begin{figure} 
		\centering
		\includegraphics[width=0.99\linewidth,keepaspectratio]{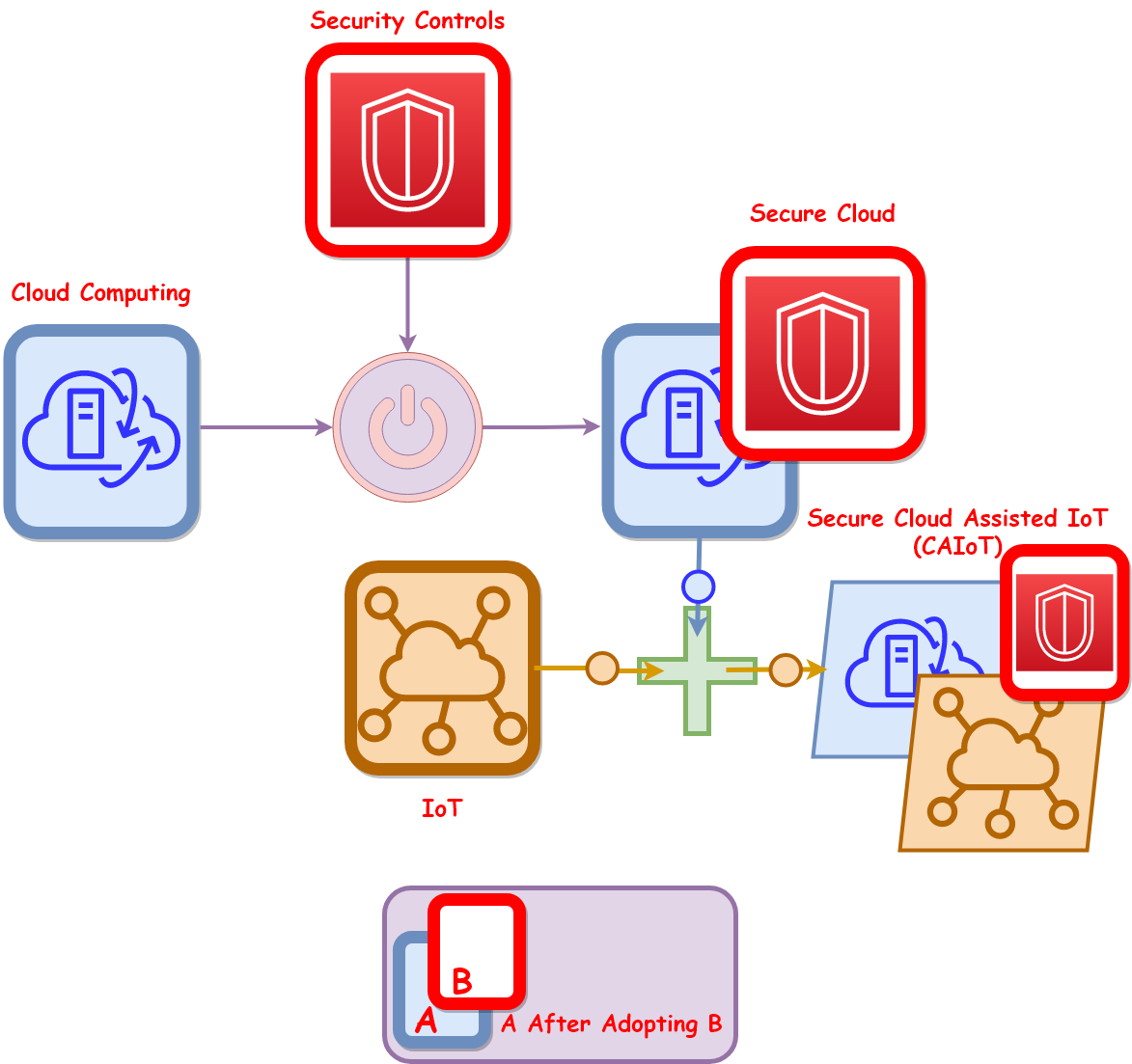}
		\caption{\textcolor{black}{IoT on Top of Secure Cloud}}
		\label{fig:SecureCloudforIoT}
	\end{figure}

	\textcolor{black}{Figure~\ref{fig:SecureCloudforIoT} shows the notion of IoT on top of secure Cloud. Security of IoT-aware Cloud has been studied by researchers in terms of the following challenges and controls.}
	
	\subsubsection{\textcolor{black}{Security Challenges}}
	In this subsection, we review existing research works focusing on different security challenges in CAIoT including privacy, trust and attack resilience. 
	
	\paragraph{\textcolor{black}{Privacy}}\mbox{}
	
	Privacy as a security challenge in CAIoT has been studied in~\cite{Chall-Conf036, b1, b2}. The authors propose a three-party model integrated with two-stage decryption, namely Paillier based cryptosystem, that allows cloud servers to interact securely along with Machine Learning Service Providers~(MLSP) by encrypting the data. The tensor-based multiple clustering methods can be widely used in the Industrial Internet of Things (IoT) to improve productivity and quality. Privacy-preserving tensor-based multiple clustering methods on the secure hybrid cloud were proposed in~\cite{Chall-Jour039}. This method employs homomorphic cryptosystems to encrypt object tensors and then uses servers to apply multiple cluster calculations upon encrypted object tensors. In addition, security subprotocols are provided to support multiple clustering methods while considering privacy issues. In similar research work~\cite{Chall-Jour046} the challenges of privacy-preserving analytics in healthcare information systems were discussed. This study is based on kHealth—a healthcare information system developed and employed for disease monitoring.
	
	The research work in~\cite{Chall-Jour046} was extended in~\cite{Chall-Jour044} to introduce a Symmetric Key InterCloud Authentication and redeemable Micropayment protocol (SKICAP) for IoT applications. This protocol ensures that authentication for mobile IoT devices is handled by edge nodes outside of their home cloud coverage. It also uses a cryptographic hash function to protect the device from privacy breaches and is resilient to security compromise.
	
	\paragraph{\textcolor{black}{Trust}}\mbox{}
	
	In~\cite{Chall-Conf045}, a 3-tier cloud-cloudlet-device hierarchical trust management protocol, called IoT-HiTrust, for IoT systems was presented. This protocols allows an IoT device to report its experiences and determines the trueness of another IoT device for service composition by following a simple localized report-and-query paradigm.
	
	\paragraph{\textcolor{black}{Attack Resilience}}\mbox{}
	
	\textcolor{black}{Regarding attack resilience, blockchain proves to be a sustainable solution in terms of flexibility, robustness, and confidentiality. Thus t}he authors of~\cite{Chall-Jour045} proposed a blockchain-enabled distributed security framework using edge cloud and software-defined networking (SDN)\cite{c6}. The SDN-enabled gateway offers early detection of security attacks by identifying and blocking suspicious network traffic or activity. Experimental evaluation results prove that this approach is efficient against various security attacks. A similar framework was presented in~\cite{Chall-Conf043} to alleviate security attacks using the Block-SDoTCloud architecture to strengthen security in the cloud storage network. To equip the cloud storage environment with the required facilities, two leading technologies, i.e., SDN and Blockchain, were applied.

	As an extension of ~\cite{Chall-Jour045}, the authors of~\cite{Chall-Conf044} proposed a blockchain-based system that tags a unique device identifier and uses smart contracts to store data transaction records in the blockchain\textcolor{black}{. This is done} for each IoT device as well as cloud and user access. It provides distributed records that can mitigate data tampering and data snitching attacks. \textcolor{black}{ The presented scheme makes use of two different methods to secure the privacy of IoT data. The first method used was a Merkle Tree data structure, which is used to efficiently and securely encode data. The second method used was Practical Byzantine Fault Tolerance (pBFT) that allows the system to continue working if some nodes become corrupt or fail. These methods are implemented into a blockchain system for secure IoT data transmission.}
	
	\subsubsection{\textcolor{black}{Security Controls}}
	
	\paragraph{\textcolor{black}{Authentication}}\mbox{}
	
	\textcolor{black}{Due} to the continuously raising security challenges \cite{b3}, the authors of~\cite{Chall-Conf041} designed a security model for Cloud-Based IoT. This model is based on blockchain technology. It uses identity verification algorithms to ensure that the data is securely stored in the cloud and transmitted from the cloud to the users. \textcolor{black}{The authors of \cite{Chall-Conf041} designed this scheme to help ensure the trustworthiness of data, low cost of interactions while providing a high level of security, and prevent tampering of data. Each of these help to create a high level of authenticity for data.}
	
	\paragraph{\textcolor{black}{Encryption}}\mbox{}
	
	Cryptography is the art and science of leveraging algorithms, mathematical problems and structures, secret keys, and complex transforms to keep data confidential while being stored or transmitted. Modern cryptography is connected with several branches of science and technology including, but not limited to quantum computing \cite{Acc-Jour-New001}, information theory \cite{New-Entropy-Jour001}\cite{Entropy-Jour001}, chaos theory \cite{ASI-Chaot-Enc} and  hardware technology \cite{Acc-Book2} in its ecosystem. The adoption of Cryptography has played a critical role in the evolution of AI \cite{ASI-Ware-Peace} in a way that it cannot be separated from AI anymore.

	\textcolor{black}{Using Mystiko Blockchain (which uses a novel smart contract platform to allow for more throughput of data) \cite{bandara2018mystiko}}, the authors of~\cite{Chall-Conf034} proposed a lightweight, decentralized, encrypted cloud storage architecture called Yugala. The Yugala architecture ensures file confidentiality, eliminates data deduplication, and increases file integrity. File confidentiality with data deduplication is based on two approaches, one is based on \textcolor{black}{convergence encryption and double hashing while} the other approach is based on symmetric encryption. This problem was also studied in~\cite{Chall-Jour038} where an optimization process was proposed. This process considers two critical aspects, the new cloud edge on-demand service model for allocation of resources and the impact of the application implemented in terms of cost, performance, and security. 
	
	The author of~\cite{Chall-Jour043} proposed a flexible, economic model by combining fog computing and cloud computing for addressing challenges in data processing, effective data retrieval, secure data storage \textcolor{black}{and} dynamic collection of data. Cloud-based IoT requires large space to store data in the cloud.  In response to this, the authors of~\cite{Chall-Conf038} proposed a framework to store data using various compression techniques and used the security standard AES encryption to secure the data. \textcolor{black}{The scheme is broken up into three different layers. The first layer is a sensor network which is responsible for collecting the data and then encrypted and compressed for transmission to the cloud. The next layer is cloud storage where the encrypted data is transmitted to. Lastly the client framework is used to decrypt and decompress the data the client is pulling from the cloud storage.}
	
	
	
	In another research work, the author of~\cite{Chall-Conf042} proposed a health monitoring system that is connected to an IoT system to transmit data to a local server. In case of any abnormalities in a patient’s health, the doctors are alerted via a message or email. In this case, the data transmission must be highly secured. Shamir's sharing key algorithm is used to serve this purpose, where the key and the data are split and retrieved using threshold cryptography. For heart diagnosis, a Support Vector Machine (SVM) is used for prediction. The author of~\cite{Chall-Jour042} proposed a similar secure parallel outsourcing scheme to increase the feasibility of the resource-bound modular exponentiation operation (widely used in cryptographic computations) for application in resource-constraint IoT devices. Additionally, an extension scheme for RSA is designed in this research in order to provide extra security for IoT devices. 
	
	In~\cite{Chall-Conf039}, the authors presented a new variant of RSA called the Memory Efficient Multi Key generation scheme. The \textcolor{black}{exchange} of sensitive information takes place from Cloud to IoT and from IoT to IoT devices. This approach reuses RSA with a diophantine form of the non-linear equation for memory efficiency. \textcolor{black}{The approach exhibited a high performance in evaluations} and it did not record extensive use of Euclid’s algorithm. Similar research work was conducted by~\cite{Chall-Conf040} to enforce higher security on sensitive data. Cryptography plays a vital role in securing the data. Thus, to enhance the security of data, two-hybrid cryptography algorithms were proposed. \textcolor{black}{The} RSA algorithm and DES algorithm are used to encrypt the data before storing it in the cloud. 

	\paragraph{\textcolor{black}{Access Control}}\mbox{}
	
	Working toward secure cloud for IoT, the authors of~\cite{Chall-Jour036} focused their efforts on \textcolor{black}{making} the majority of the outsourced data rely on cloud services, which increases the number of attack points as a \textcolor{black}{side} effect. \textcolor{black}{The authors address the issue by leveraging Ciphertext-Policy Attribute-Based Encryption (CP-ABE). This allowed for credentials to be assigned to users to access encrypted data. CP-ABE has the issue of incorrect protection of keys and allowing unauthorized users to access the data. These issues motivate the authors to create a more verifiable access control scheme. To do this they take advantage of blockchain as a tamper-proof log that allows for transparency and traceability of data and modifications. Access is given to data through the use of smart contracts and then the details of use is turned into blocks on the blockchain. Extensive testing completed by the authors has shown the scheme to be efficient in terms of resource use, speed, and security.}

	
	
	

	\subsection{\textcolor{black}{Secure IoT on Top of Cloud}}\label{Appr-Sub2}
\begin{figure} 
		\centering
		\includegraphics[width=0.99\linewidth,keepaspectratio]{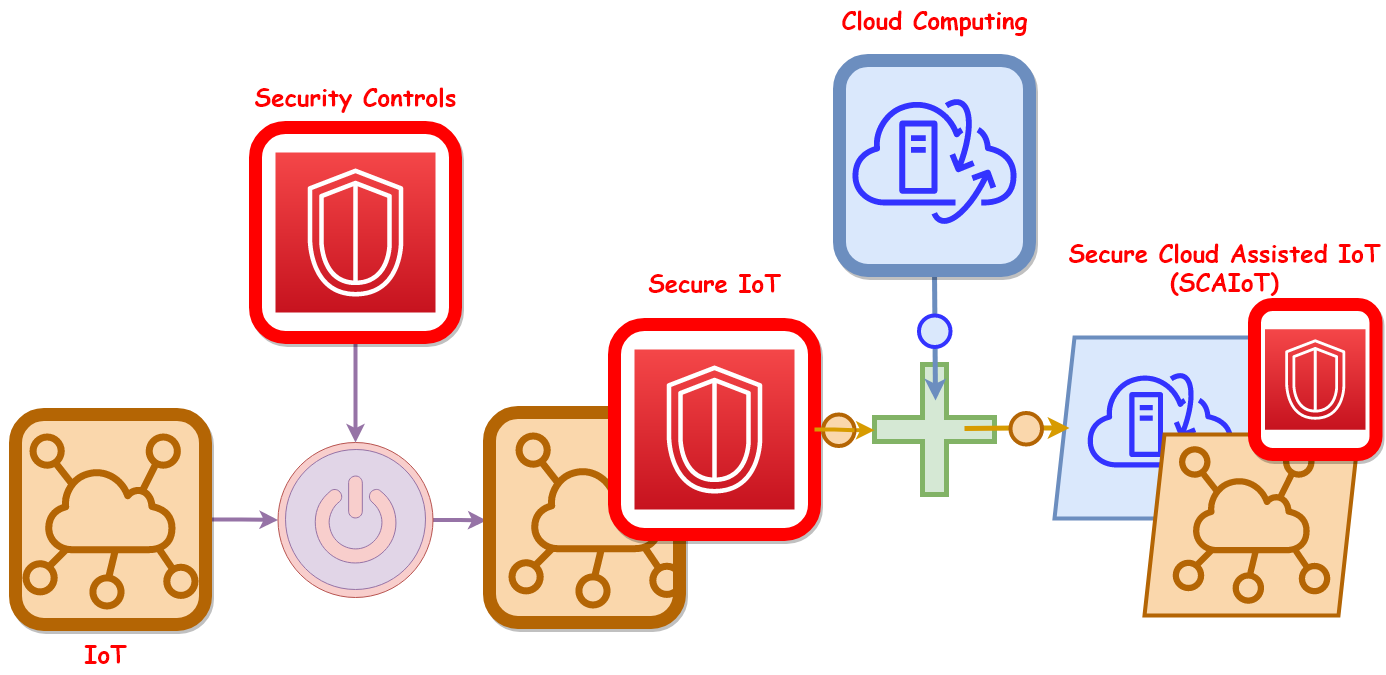}
		\caption{\textcolor{black}{Secure IoT on Top of Cloud}}
		\label{fign}
	\end{figure}

	The idea behind secure Cloud-aware IoT is illustrated in Figure~\ref{fign}. \textcolor{black}{In the following subsections, challenges and controls are studied with respect to this approach.}

	\subsubsection{Security Challenges}
	
	\paragraph{\textcolor{black}{Privacy}}\mbox{}
	
	\textcolor{black}{When focusing on privacy as a security challenge for IoT, it is of critical importance to consider an identity protection scheme. This} is shown in \cite{Chall-Conf052} \textcolor{black}{and is used for} symmetric data encryption. \textcolor{black}{The main objective of the authors is to create a novel privacy preserving symmetric encryption method that can take advantage of symmetric encryption's fast processing while preserving the data privacy. The authors take advantage of two layers the outer layer uses k-anonymity based symmetric encryption. The inner layer uses random encryption to further improve privacy and security.}\textcolor{black}{Along with the proposed scheme, the authors give security definitions for the problem. This provides security levels for the newly constructed models.} \textcolor{black}{Security analysis done by the authors shows that the proposed scheme is reasonably secure while remaining energy efficient.}
	
	This problem is further discussed in~\cite{Chall-Conf055}, where the authors propose a sensor cloud architecture that adopts security considerations as essential design objectives. It is based on a multi-layer client-server model that separates physical and virtual instances of the sensors to achieve enhanced security, privacy, and reliability. The main concept behind these virtual instances can be secured quickly. It \textcolor{black}{can also} be implemented in real-time. This motivates a lightweight data privacy scheme proposed in~\cite{Chall-Conf057} that outsources data from low-power IoT devices to the mobile cloud computing system.
	
	\paragraph{\textcolor{black}{Trust}}\mbox{}
	
	Trust, as a security challenge \textcolor{black}{in} CAIoT, has been studied in a research reported in ~\cite{Chall-Jour047}. The architecture proposed in ~\cite{Chall-Jour047} relies on \textcolor{black}{an edge network and an edge platform} to reduce resource consumption and guarantee the trust evaluation mechanism, respectively. Moreover, a data-centric approach for IoT is introduced  ~\cite{Chall-Jour047}, where edge and fog clouds leverage data-centric networks and help in the optimal handling of upstream data flow. \textcolor{black}{Each IoT device calculates trust and transmits it to the edge server for mapping. The scheme uses a trust evaluation mechanism and service template to take advantage of both of the edge and fog computing capabilities. Testing done by the authors show it is very secure and efficiently uses resources.}
	
	\subsubsection{\textcolor{black}{Security Controls}}
	
	\paragraph{\textcolor{black}{Authentication}}\mbox{}
	Recently, blockchain has been advocated as a suitable authentication mechanism \cite{b5, b4}. The framework proposed in~\cite{Chall-Conf048} uses blockchain for authentication at both the edge and cloud layers. Proof of Authority (POA) is deployed as to enhance the overall performance of this system. 
	
	Biometric authentication is widely used in various applications today. But biometric authentication using IoT devices is challenging. To handle this, an IoT system using biometric authentication over a cloud-based IoT service is described in~\cite{Chall-Conf049}. In this architecture, the system employs an IoT service platform\textcolor{black}{, NETPIE, which is used to facilitate the interconnected biometrics on IoT devices. To improve this further the authors suggest the use of DAEDALUS which messages NETPIE when malware is present. The system acts as a secure alerting mechanism. }

	\textcolor{black}{The amount of land required to produce enough food for large populations is becoming overwhelming. The use of IoT can help to maximize production of food on the land already designated for crops by using different hydroponic systems.} The authors of~\cite{Chall-Conf050} discuss the design and development of \textcolor{black}{an} automatic monitoring of an indoor vertical hydroponic system. There are three main components in this system, these are: an indoor hydroponic module, a mechanism for monitoring and handling the operation of the system, and a two-factor authentication \textcolor{black}{(2FA)} to strengthen the security elements of the system. \textcolor{black}{The authentication system is designed to first have the user login using a user login, a project login, and finally an access code to a third party system. The third party system then sends a notification to the registered users phone to further be authenticated. This allows for an additional layer of security in these sensitive IoT networks. }
	
	In~\cite{Chall-Jour050}, a lightweight authentication protocol for IoT devices, called light edge, is proposed. It consists of a three-layer scheme, the IoT device layer, trust center at the edge layer, and cloud service providers. Evaluation results show that the proposed protocol is highly effective against a number of security and privacy attacks. In~\cite{Chall-Conf054}, the authors proposed to reuse the existing WIFI infrastructure to connect the devices to the Internet. They present a tunneled-based protocol EAP (TEAP) protocol for authorizing IoT devices from the cloud using the EAP-AKA protocol as an inner method EAP.
	
	\paragraph{\textcolor{black}{Encryption}}\mbox{}
	
	The authors of~\cite{Chall-Conf047} devised a secure, low-cost device that is employed to detect forced entry via IR\textcolor{black}{. In their proposal, there is a software unlocking method before the physical unlocking} system that exchanges information in an encrypted form through a smartphone application over a secure line. 
	
	\textcolor{black}{Encryption can be very costly for systems with limited computational resources such as in IoT. The authors of~\cite{Chall-Conf051} proposed a lightweight }model that provides sensitive data access from IoT owners only to authorized user. \textcolor{black}{To do this the IoT device will send the data to a semi-trusted encryption server to encrypt the data. Once encrypted the data is stored in the cloud to be accessed by data owners.} Here, the IoT owner can manage the authorization with a single username/password to access their sensitive data. Preliminary evaluation results show that this model offers efficient key management for IoT owners.
	
	Another approach to encrypt personal information was proposed in~\cite{Chall-Conf059}. Zero-Knowledge proof blockchain solution is implemented using block inquiry to protect personal information against security threats. In~\cite{Chall-Jour052} AES is proposed is proposed for signal encryption and decryption in electrocardiogram instruments. However, this method does not perform well with low-power IoT devices. 
	
	The research reported in~\cite{Chall-Jour051} has led to the design of a model that takes a computational approach towards secure key generation in order to safeguard the data in CAIoT using encryption. The proposed key generation scheme consists of the following components.
	
	\begin{itemize}
		\item In the first component, an encryption service model based on Support Vector Machine (SVM) is run using a key generation from the conventional encryption operation mode with some improvements.
		
		\item Optimization techniques are added to the key generation process to make it computationally more secure, specifically for /cloud environment.

	\end{itemize}
	
	\paragraph{\textcolor{black}{Access Control}}\mbox{}
	
	Nowadays, access to cloud services is rendering a bottleneck in single-Cloud infrastructures due to the increasing demand. To alleviate this bottleneck, the authors of~\cite{Chall-Jour049} have proposed a scheme that provides access to the best IoT services among other services in a multi-cloud IoT. 
	This scheme consists of the following three stages of operation.
	\begin{itemize}
		\item Choosing the best service among available IoT services.
		\item An authentication stage aimed at authenticating users during execution.
		\item Service-Level Agreement (SLA) management.
	\end{itemize}
	
	\paragraph{\textcolor{black}{Reputation Management}}\mbox{}
	
	Data leakage and privacy exposure are considered as serious threats \textcolor{black}{these days}. To respond to this challenge, the author of~\cite{Chall-Jour048}  proposed a \textcolor{black}{reputation evaluation system. The scheme uses S-AlexNet convolutional neural network and dynamic game theory to determine a secuirty reputation for a health care data system. This encourages system administrators to provide a high level of privacy for health data. Furthermore the scores can then be used to protect IoT systems against privacy attacks.} 
	
	\paragraph{\textcolor{black}{Intrusion Detection}}\mbox{}
	
	\textcolor{black}{Intrusion detection is important in any system including CBIoT. The authors of \cite{Chall-Conf056} first break down the gaps of IoT embedded software and physical devices of IoT as well as the security flaws that are attached to them. To bridge the gaps the authors make a disconnect between IoT functions and IoT devices by making a clone of IoT devices on the cloud. These clones are stored on an IoT Agent Platform and can be used to simulate function calls to physical IoT devices. To allow for transparency of the devices the authors implement Dripcast, a programming framework to allow physical devices to access the clones stored on the cloud IoT Agent Platform. This allows easy development of IoT devices as designers can transparently see how they interact with the clones. This scheme can allow for more efficient and secure designs of IoT devices and functions preventing further intrusions on a system.}
	
	
	\paragraph{\textcolor{black}{Risk Assessment}}\mbox{}
	
	To analyze the security risks, another research advancement, presented in~\cite{Chall-Conf053}, uses attack tree analysis, soft systems thinking, and red team techniques to minimise  cybersecurity risks. Using the information security research lens, it was possible to anticipate real-world attacks using trained data and assess the risks from a cloud perspective.
	
	
	\subsection{\textcolor{black}{Secure IoT-Cloud Communications}}\label{Appr-Sub3}
	
\begin{figure} 
		\centering
		\includegraphics[width=0.99\linewidth,keepaspectratio]{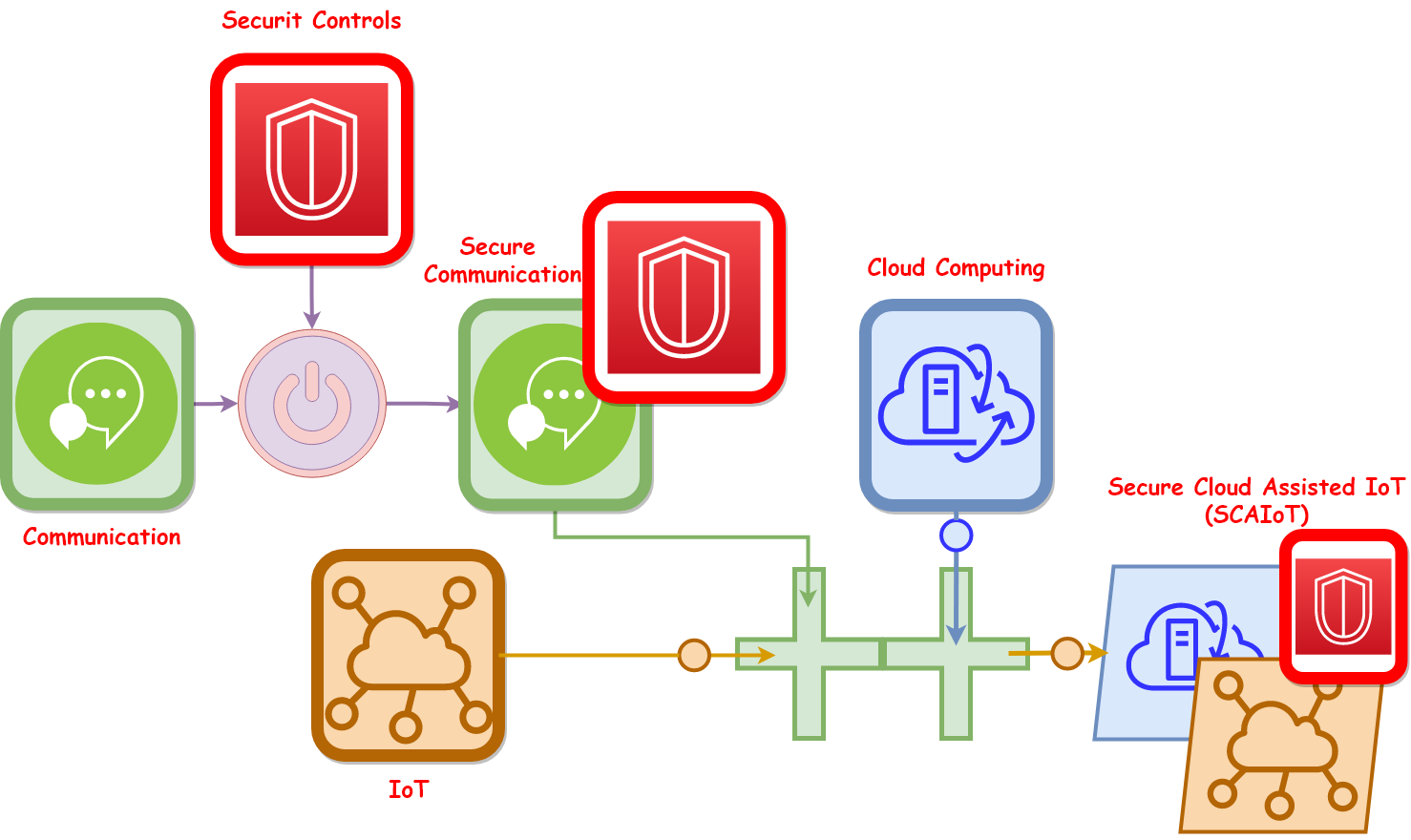}
		\caption{\textcolor{black}{Secure IoT-Cloud Communications}}
		\label{fig:SecureIoTCloudcommunication}
	\end{figure}

	This approach is shown in Figure \textcolor{black}{\ref{fig:SecureIoTCloudcommunication}}. Research on this approach has led to the study of the security challenges in the following subsections.
	\ref{fig:SecureIoTCloudcommunication}.		

	\subsubsection{\textcolor{black}{Security Controls}}
	
	\paragraph{\textcolor{black}{Authentication}}\mbox{}
	
	IoT devices are susceptible to various security attacks \cite{b6, b7}. To prevent their success, the authors of~\cite{Chall-Conf060} proposed a design by integrating the IoT module with the Amazon IoT platform and securing access using authentication. \textcolor{black}{The authentication for new IoT devices is done via certificate signing requests (CSR). The CSR requires certain information including a key to be sent to the AWS platform so that it may be authenticated. The Amazon server returns a certificate to the IoT device that can be used for mutual authentication of the IoT device and the Amazon IoT platform. These methods are able to then prevent attackers from accessing data or communicating with the IoT devices or the Amazon IoT platform.}
	
	
	\paragraph{\textcolor{black}{Security Evaluation}}\mbox{}
	
	Security evaluation is one of the most important security challenges \cite{b9}. This problem was discussed in~\cite{Chall-Conf061} where the authors proposed Inspection-Friendly Transport Layer Security (IF-TLS). \textcolor{black}{This scheme is used as an  interception-friendly communication library. Unlike other implementations of TLS in IoT IF-TLS can be fully controlled by residential gateways. Along with using TLS for encryption the scheme uses middleboxes to allow the user to analyze traffic. This scheme is an introduction of what can be done using IF-TLS as the authors believe a lot can be done to improve the proposed scheme.}

	
	\subsection{Integrated \textcolor{black}{SCAIoT} Solutions}\label{Appr-Sub4}
	
\begin{figure} 
		\centering
		\includegraphics[width=0.95\linewidth,keepaspectratio]{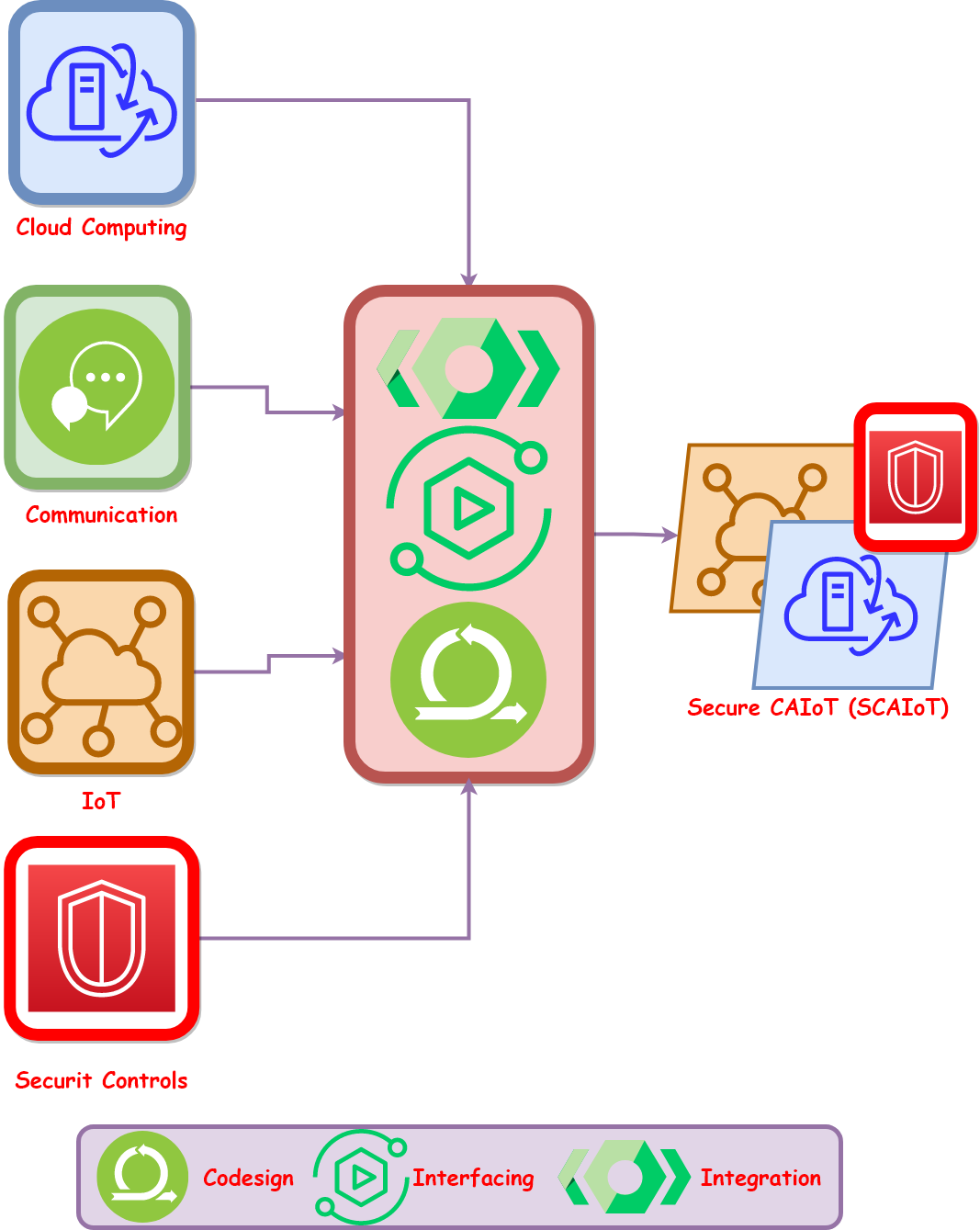}
		\caption{\textcolor{black}{Secure Codesign, Interfacing and Integration}}
		\label{fig:IntegratedSolution}
	\end{figure}

	Figure~\ref{fig:IntegratedSolution} visualizes an integrated solution for the design of secure CAIoT. The following challenges, \textcolor{black}{mechanisms} \textcolor{black}{ and controls} have been studied in research works focusing on this approach.
	
	\subsubsection{\textcolor{black}{Security Challenges}}
	
	\paragraph{\textcolor{black}{Privacy}}\mbox{}
	To handle the privacy challenge \cite{c7}, a new network attestation scheme that provides scalability \cite{b8}, forward security, and preservation of privacy simultaneously was presented in~\cite{Chall-Jour004}. This approach ensures the confidentiality of previous messages even when the current secret key is revealed. 
	Existing cloud security models employ Symmetric Search Encryption~(SSE) to search for encrypted data. However, this search activity requires the data owner to share the real key with query users by ignoring malicious cloud servers. These limitations pose a major challenge when implementing SSE. To overcome these limitations, a framework was introduced in~\cite{Chall-Jour002} for cloud-based IoT. It initially provides a protocol with limited key disclosure without exposing plaintext query points. This technique processes the search using Merkle hash tree structure and k-means clustering. This approach is secure against level-2 attacks. For level-3 attacks, the random splitting technique is used. 
	
	Due to the openness of the cloud, retrieval of data from the cloud will lead to security and privacy issues. To alleviate this, Privacy Enhanced Retrieval Technology~(PERT) was proposed in~\cite{Chall-Jour003} for Cloud-based IoT. PERT ensures privacy by hiding the information between the edge and cloud servers during data transmission. In this approach, privacy is guaranteed because edge servers maintain partial data, and the attacker has to maintain the data between both edge and cloud servers.

	Cloud-based IoT consists of numerous innovative systems monitored and controlled via the Internet. In this model, wireless communication is considered to be the primary source of energy usage. It can be avoided by limiting radio communications. However, privacy issues pose a significant challenge for these types of devices. Hence, to address high energy usage, an Improved Ant Colony Optimization (IACO) is presented in~\cite{Chall-Jour012}. A privacy-preserving method is also proposed for privacy concerns. The published experimental results prove that the proposed way is efficient in terms of network lifespan and energy consumption.
	
	Based on the research work in~\cite{Chall-Jour021}, the data encrypted should maintain privacy when stored in IoT cloud storage. The author developed a cipher-text-policy attribute-based encryption scheme, which assigns access control to the cloud. With the help of this model, the data is secured against the plaintext attack.

	\textcolor{black}{Data sharing happens frequently in the internet of medical things (IoMT). The issue is that many devices are resource constrained so encrypting that data is not always possible using traditional methods.} In~\cite{Chall-Jour025}, the  proposed framework suggests using CVaR to identify risk based on channeling delay. The authors optimally developed a branch-and-check (BNC) approach with minimum cost and latency. Based on the results obtained, the authors advocate that the unnecessary operation cost can be minimized and optimized \textcolor{black}{to provide secure transmission of data}.
	
	Next, Sharma, et al.~\cite{Chall-Jour027} proposed an LABSE scheme and a model that is IoT-based on cloud medical applications. The system is beneficial in real-time scenarios where the security and privacy of patient data are secured. In~\cite{Chall-Conf004}, the authors proposed a secure system for data privacy that helps in accessing the data which is in an encrypted format. The authors match the attributes and collect the partial attributes with several advantages. To \textcolor{black}{meet privacy-preserving requirements}, the authors of~\cite{Chall-Conf019} suggested an encryption approach with the keyword, which supports a multi-user environment with the inverted index for Cloud-IoT Platforms. 
	
	In \cite{Chall-Conf023}, the first-ever accountable and revocable large universe Multi-Authority with decryption based on additive groups in Edge-Cloud IoT is presented. This scheme is fixed securely in the oracle model under the q-DPBDHE2  assumption, providing service quality Loss-aware privacy protection. Similarly, in~\cite{Chall-Conf028} the authors present a concept of a data bank to secure user sensitive data by controlling the devices and contains several layers from IoT objects to web and mobile devices in the top layer. The authors of~\cite{Chall-Conf029} \textcolor{black}{focused} on the privacy and security of cloud-based IoT environments. They addressed analyzing and comparing current approaches in ensuring fundamental security requisites and securing intercommunication of IoT.

	\textcolor{black}{Processing large amounts of data is extremely important in the realm of IoT. Due to its ability to efficiently and with low latency transmit data, cloud edge computing has been studied vastly in recent years. The authors of~\cite{Chall-Jour041} present a scheme of edge cloud computing that takes advantage of PUFs while solving some of the issues related to the synchronization of challenge response pairs (CPR). The scheme uses blockchain to provide a more synchronized scheme to allow for a secure and efficient form of cloud edge computing for IoT data transmission. The random oracle model is used in testing the scheme and proves its security.}
	
	
	\paragraph{\textcolor{black}{Trust}}\mbox{}
	
	Trust is studied using the framework discussed in~\cite{Chall-Jour017}\textcolor{black}{. This scheme} evaluates the trustworthiness of cloud services to conform to the security standards of the cloud-based IoT by integrating security and reputation-based trust assessment methods. The security-based trust evaluation method utilizes cloud-specific security metrics to evaluate the security of a cloud service. Furthermore, the feedback on the quality of cloud service is reviewed to evaluate the reputation of a cloud service. 
	
	Also, the authentication system was evaluated, and its flaws were demonstrated by Alzahrani, et al.~\cite{Chall-Jour030}. Then, in comparison to the related systems, the authors suggested a safe and lightweight authentication and key agreement scheme that meets fundamental security criteria while incurring the little computational cost.
	
	In the research carried out in~\cite{Chall-Conf024}, the author has suggested the scheme helps user-attribute revocation. The proposed scheme \textcolor{black}{uses a MAPE-K feedback control loop to determine a trust ranking for an IoT cloud system. This allows system administrators to make appropriate changes to provide} support and security against attacks. According to the analysis methods results, it is ideal for large-scale cross-domain collaboration in the dynamic cloud-aided IoT. \textcolor{black}{ The authors of~\cite{Chall-Conf031} propose} a game-theoretic approach for recruiting trustable Sensing service Providers (SSPs) in mobile crowd sensing systems.
	
	\paragraph{\textcolor{black}{Integrity}}\mbox{}
	
	The authors of~\cite{Chall-Jour035} research and create a formal access control model for Google Cloud Platform, known as the GCPAC model. They next develop the GCPAC model into a formal Google Cloud Platform IoT Access Control (GCP-IoTAC) model with IoT-specific components. Using the GCPIoT, they illustrate two popular IoT scenarios, E-health and smart home. They also highlight some of the limits of GCP-present IoT's access control capabilities and propose attribute-based additions for fine-grained access control in GCP and its IoT platform. 
	
	Finally,~\cite{Chall-Jour041-1} \textcolor{black}{proposed a data auditing scheme where third party auditors audit data integrity. The authors then provided a secondary layer for auditing the third party auditors. The scheme uses a cloud-stored auditing method for IoT devices based on consortium blockchain. This} allows IoT device data owners to analyse auditor's activities. \textcolor{black}{Testing of the system proves an increase in data integrity and security in using third party auditors.}
	
	\paragraph{\textcolor{black}{Attack Resilience}}\mbox{}
	
	To address attack resilience problem, the author of \cite{Chall-Jour011} proposed a user-centric model, ARCA-IoT. It is defined as a system that determines the attributes essential for trust and provides a user-centric model capable of tackling attacks wisely. For scalability and interoperability, a cloud-based environment is introduced in the ARCA-IoT. 
	An instinctive Naive Bayes approach is employed to train this system so that it manipulates the possibilities of the trustworthiness of the entities and then determines different types of attacks with the help of algorithms. Cloud-assisted IoT is more susceptible to malware attacks. Thus, in order to maintain confidentiality and integrity, efficient encryption mechanisms must be implemented to protect the data against different kinds of security attacks. Despite this, it is not easy to perform calculations on data once it is encrypted. Fully homomorphic encryption can be implemented in this regard, but it is inefficient in terms of data manipulation. However, with the help of a semi-trusted server, this type of encryption can be used, but the issue behind this is it cannot be used for multiple distributed systems with shared IoT data.

	\subsubsection{\textcolor{black}{Security Controls}}
	
	\paragraph{\textcolor{black}{Authentication}}\mbox{}
	
	Authentication is a crucial issue since biometric data is difficult to counter and replace if it is compromised. To overcome this issue, an approach based on fuzzy commitment protocol, a fingerprint recognition scheme using minutiae-based sector coding strategy is presented in~\cite{Chall-Jour009}. In this approach, the minutiae of the fingerprint are classified into various designed sectors and then encoded based on their features. Furthermore, numerical results were presented to prove that this method is practical.
	
	\textcolor{black}{Many identity management schemes are costly to implement in IoT platforms. To solve this}, Ahmad, et al.~\cite{Chall-Conf005} \textcolor{black}{created a decentralized blockchain network that takes advantage of facial recognition for identification. The cloud is used to perform complex tasks in the scheme that couldn't be done on most IoT devices. Once the cloud gathers facial ID date from the user, the blockchain confirms the identity of a user and creates a smart contract. The smart contract is used to facilitate the terms of use for the service. This is then stored on the blockchain. The use of blockchain and the cloud reduces the stress on the IoT devices and increases transparency, which improves security of the overall system.}

	
	\textcolor{black}{The research done in} ~\cite{Chall-Conf010} describes a technique and gives a tutorial on IoT authentication, laying the groundwork for future research into protocol validation. For IoT-enabled smart devices, an extremely lightweight mutual authentication mechanism based on Diffie Hellman has been designed. The author \textcolor{black}{of} ~\cite{Chall-Conf012} gave a classification of likely occurrences, proposed a knowledge-based approach, and conducted a rigorous analytic investigation.	
	
	In another study, Alsahlani, et al.~\cite{Chall-Conf015} \textcolor{black}{has the main focus of improving Wazid et al.’s \cite{wazid2020lam} cloud-based authentication system. The authors suggest the removal of multiple smart cards to be replaced by a single one. It is also suggested to remove the clustering strategy proposed in Wazid et al. to allow for greater scalability. Lastly the authors of ~\cite{Chall-Conf015} suggest the use of a suitable authentication method to prevent data leakage.}
	
	

	In~\cite{Chall-Conf016}, a lightweight two-factor mutual authentication scheme for Cloud-based IoT is proposed. Also, the authors have commented on the weakness of state-of-the-art methods. Moreover, in~\cite{Chall-Conf020}, the authors suggested an architecture known as data bank. 
	In~\cite{Chall-Conf021}, a safe and efficient KNN query has been proposed to be applied on encrypted untrustworthy data. The authors utilized modified homomorphic encryption to encrypt the object set, which allows for addition and multiplication over encrypted data. Because the anticipated rank meets the key top-k characteristics, it may be a confidence ranking criterion.
	
	A lightweight authentication solution for cloud-centric IoT applications was proposed in~\cite{Chall-Conf032}. The scheme's official security verification was carried out utilising the widely used AVISPA tool. Furthermore, the security analysis, computational and communication costs show that the proposed lightweight system is secure and efficient. \textcolor{black}{Another important area is} secure multi-factor remote user authentication \textcolor{black}{schemes. The} authors of \cite{Chall-Conf037} examined the security of a state-of-the-art method in their research. \textcolor{black}{They demonstrated} that the method is vulnerable to identity and password guessing, replay, and session key disclosure attacks. To address the security issues in the evaluated method, they also presented a safe multifactor authentication system for cloud-IoT environments.

	\paragraph{\textcolor{black}{Encryption}}\mbox{}
	
	\textcolor{black}{Attribute-Based} Encryption (ABE) requires predefined public parameters during system initialization \textcolor{black}{including} performing encryption and decryption efficiently, and achieving security by assuming basic security standards. To address the challenges present in the existing models, an unbounded and efficient attribute-based encryption scheme with compatible security for cloud-based IoT is presented in~\cite{Chall-Jour001}. Compared to the previous approaches in this field, this scheme not only provides access control over encrypted data in a fine-grained way\textcolor{black}{, but it has also been proven} to be secure under standard decision linear assumption. Moreover, it \textcolor{black}{eliminates} the need for the initialization of predefined public parameters.
	
	To ensure the normal operation of the system in IoT devices, remote attestation for every smart device should be ensured. \textcolor{black}{Due} to the engagement of large numbers of smart devices, traditional attestation schemes are unable to meet the requirements in terms of efficiency. Security risks due to secret key exposures are more frequent in traditional schemes \cite{c1}. To address this issue, a new method with the help of a semi-trusted server is introduced  in \cite{Chall-Jour013}. This \textcolor{black}{method is} capable of performing manipulations on the encrypted data without gaining any useful information. A fine-grained self-controlled outsourced data deletion scheme in cloud-based IoT is presented in \cite{Chall-Jour015}. The main contribution of this scheme is that it enables data owners to precisely and permanently delete their outsourced IoT data in a policy-based way without depending on the Cloud Server. 
	
	The authors of~\cite{Chall-Jour005} proposed an effective data sharing technique that engages intelligent devices to share data securely at the edge of cloud-based IoT. A secure searching scheme to search for the particular data is also proposed. Retrieval of Ciphertexts under heterogeneous datasets is of utmost importance in Cloud-based IoT devices used today. Existing ciphertext retrieval schemes cannot retrieve ciphertexts from encrypted heterogeneous datasets. To alleviate this problem, the authors of~\cite{Chall-Jour006} proposed a transparent ciphertext retrieval system without considering the programming language, accessible platforms, and heterogeneity of the datasets. This system acts as a middleware with an encrypted cloud database system. The middleware facilitates cross-language and cross-platform queries over different database systems, and it also realizes cross-database queries over the encrypted cloud datasets. Cloud-assisted IoT is regarded as an emerging technological trend in the IoT domain, where the cloud acts as a great storage provider for storing a vast amount of data.
	
	To ensure the confidentiality of the data, the encryption mechanism is designed so that the user assigned by the data owner can only decrypt the data. However, there is a requirement in a multi-cloud IoT environment, and the data has to be shared with multiple users rather than the designated user. To satisfy this, the authors of~\cite{Chall-Jour007} proposed a scheme Flexible privacy-preserving data sharing (FDPS), where the user can encrypt the data to a shared user using identity-based encryption. The user can employ a fine-grained access policy to generate a credential and share it. This, in turn, converts that credential into new ciphertexts that can be readable by the shared user.
	
	CAIoT provides a great solution to growing data size problems for the constraints of separate objects. However, with the implementation of the cloud, IoT faces new security challenges for data exchange, which are introduced in this paper and not addressed in traditional approaches. The authors of \cite{Chall-Jour008} envisage a secure cloud-assisted IoT to protect data confidentiality when collecting, storing, and accessing IoT data. Due to the rapid development in IoT, security and privacy challenges pose a significant threat. Biometric authentication integrated with IoT can be considered an adequate replacement compared to traditional biometric authentication.
	
	The \textcolor{black}{authors of} \cite{Chall-Conf001} proposed searching over encrypted data for multiple data senders in the cloud\textcolor{black}{. This is done through} public-key encryption with keyword search (PEKS) \textcolor{black}{and} has been regarded as one of the best \textcolor{black}{forms of} searchable encryption (SE). Unfortunately, prevailing PEKS schemes are currently susceptible to file-injection attacks due to the lack of forwarding privacy. Thus, a forward secure PEKS scheme was proposed based on identity-based encryption for cloud-based IoT environments. Searchable public-key encryption is an introductory module to realize fast keyword search upon ciphertexts present in the cloud. 
	
	
	To mitigate this problem, the authors of \cite{Chall-Jour010} proposed a new instance of SPCHS to attain fast and simultaneous keyword searches over public-key ciphertexts. It is considered \textcolor{black}{to be a} new type of relationship among searchable ciphertexts \textcolor{black}{, which} is built by the latest instance, where each searchable ciphertext is associated with a standard and public parameter.
	
	All associated relationships are disclosed on receiving a keyword search, and then all matching ciphertexts are found. Similarly, phrase search enables the retrieval of authentic documents, which is essential for machine learning applications incorporated in cloud-based IoT. To prevent sensitive data leakage from service providers, data is encrypted before being outsourced to the cloud. This leads to a big challenge in searching for encrypted data. Existing methods also do not perform exact phrase searches as they cannot determine the location relationship of multiple keywords over encrypted data. Thus, a privacy-preserving search scheme \textbf{“P3”} is discussed in \cite{Chall-Jour016}. It uses homomorphic encryption and bilinear /\textcolor{black}{mapping} to obtain the location relationship of queried keywords over encrypted data. In addition to this, it also employs a probabilistic trapdoor generation algorithm to protect users’ search patterns.
	
	According to authors~\cite{Chall-Jour019}, a phrase search scheme based on IoT data can be used as the model\textcolor{black}{. It} not only returns the file with the query, but the pending files also do not contain query phrases as well. The model stores create a tag for storing the data with the help of identifiers, then identifying with the help of identifiers, and then lastly generating a verification tag. As a result, the system is secure and efficient.
	
	As the data and cloud server interactions are increasing day by day, data privacy is a vital factor to keep privacy. The authors of \cite{Chall-Jour020} developed a scheme for querying data to cloud storage. For the ciphertext, the \textcolor{black}{authors} used the Paillier cryptosystem for encrypting the data. The \textcolor{black}{authors'} findings are secure and efficient.
	
	In~\cite{Chall-Jour026}, the authors propose a productive and fine-grained data sharing \textcolor{black}{scheme} with the help of keywords that helps in removing hostile data users. A system is designed in such a way that the user identity is unknown to provide data accuracy. 
	
	
	The authors of~\cite{Chall-Conf006}  suggested improving IDM in IoT- cloud by combining face recognition \textcolor{black}{with} blockchain-based transaction register. The validation of face recognition will allow the user to access printing services. In~\cite{Chall-Conf007} attribute-based encryption is used for ensuring the externalization data from multiple users. With the help of PU-ABE, the input text is shorter in size, and the size of ciphertext supports the access policy. Moreover, ElGamal encryption and an identity-based signature method are used to create a safe, lightweight data aggregation scheme for cloud-assisted IoT devices. 
	
	
	\textcolor{black}{The} authors of~\cite{Chall-Conf009} considered the literature survey reviewed \textcolor{black}{and} suggest that ABE is the ideal technique for cloud-based IoT. Bilinear maps were utilized in the ABE algorithm, which adds complexity and renders it unsuitable for IoT devices. Elliptic curve cryptography-based techniques have reduced processing costs \textcolor{black}{more} than bilinear mappings. Guechi, et al.~\cite{Chall-Conf017} proposed a simple two-factor mutual authentication scheme. The proposed scheme meets the requirements of mutual authentication, the security of a secret key, session key agreement, resistance to forgery, insider attack, replay attack, and impersonation attack. In~\cite{Chall-Conf018}, the authors  addressed various obstacles that multiple users in shared data face. Indeed, secure and parallel expressive Search over encrypted data with access control in \textcolor{black}{multi-Cloud IoT} has been implemented to improve availability, integrity, and privacy.
	
	Researchers in~\cite{Chall-Jour028} focused on 5G-IoT networks with SDN/NFV, blockchain-based vulnerability detection, and mitigation paradigm is provided. The five planes that make up this concept are the 5G infrastructure plane, data plane, blockchain plane, control plane, and application plane. This architecture aims to protect users from their sensitive data via Blockchain-based lightweight Security in the 5G/B5G enabled SDN/NFV cloud of IoT.

	The authors' of study~\cite{Chall-Jour029} promise to speed encryption and to search. A lightweight built edge-aided searchable public-key encryption as ESPE used outsourced public-key encryption with keyword search as PEKS and hidden structures in its models. This is a Lightweight Secure Searching \textcolor{black}{scheme} for Industrial IoT (IIoT) Devices.
	
	Researchers ~\cite{Chall-Jour031} focused on the decentralized multi-authority attribute-based encryption for Cloud-Aided IoT. CloudIoTSecurity is a quality model and technique to evaluate security in Cloud IoT applications aligned to ISO 25040, which is a help to assess security step by step towards the discovery of security concerns and, therefore, the improvement of these applications. These work to minimize the risk of attacks via protecting \textcolor{black}{ the, privacy and the reliability of the data}. 
	
	\textcolor{black}{Cloud IoT networks are dynamic and need a dynamic form of access control over the data that is encrypted in communication.} 
	In~\cite{Chall-Jour032}, the authors proposed to communicate compensation related to the authenticity of the location by workers\textcolor{black}{. This is to minimize the effectiveness of collusion attacks}. Indeed, it is a secure, efficient revocable large Universe multi-authority Attribute-Based Encryption for Cloud-Aided IoT.  \textcolor{black}{Using the random oracle model the authors were able to show the proposed scheme is a secure for of access control for encrypted data.}
	
	Another work,~\cite{Chall-Jour033}, introduces an autonomic trust management system for \textcolor{black}{Cloud-based highly} dynamic IoT applications and services. We are becoming increasingly conscious of the importance of removing the influence of a third party's malevolent recommendation and defamatory practices on evaluation findings.
	
	\textcolor{black}{With IoT having such a large amount of data it is more and more important to have a secure place to store it, such as a cloud. To keep data secure on the cloud the} research in \cite{Chall-Jour040} propose LH-ABSC, a lightweight ABSC method that uses ciphertext-policy encryption (CPABE) and key-policy attribute-based signature (KPABS). \textcolor{black}{Fog nodes are used to reduce computing overhead in the scheme. The authors provide an in-depth analysis to show the security of the system.}

	\paragraph{\textcolor{black}{Access Control}}\mbox{}
	
	To address access control \textcolor{black}{issues}, Kaaniche, et al.~\cite{Chall-Conf003} studied in detail and researched what problems \textcolor{black}{there are} in light-weighted computation. The authors proposed an offline computation that will help to retrieve the encrypted data without giving away the delicate information.
	
	The author of this research~\cite{Chall-Conf013} suggested a blockchain-based cloud-based IoT device authentication system. A functional proof of concept was created and successfully built using a hyperledger composer to test the suggested approach. This was accomplished by extensive investigation and careful evaluation of technological choices. The suggested method meets the CIA's secrecy, integrity, and availability requirements. The confidentiality requirement is met by prohibiting unauthorized access to IoT network devices. 
	
	Mheni, et al.~\cite{Chall-Conf014} presented an effective outsourced data access control mechanism for cloud-based IoT with user revocation. Not only may most revocation-related activities be outsourced to the cloud via \textcolor{black}{their} method, but the efficiency of encryption and decryption has also been improved. 
	
	Article~\cite{Chall-Conf026} provided a method for considering cybersecurity as an IR dimension for cloud-based services. They compare the related works through a compact survey and conclude with a multi-perspective debate to incorporate cybersecurity in the IoT/IoTSE architectures, based on edge powered searching over IoT from a theoretical perspective.

	\paragraph{\textcolor{black}{Security Evaluation}}\mbox{}
	
	\textcolor{black}{The cloud is a rather new invention that is being implemented into many different areas of IoT. Many people fear that it is not secure and this is why security evaluation is such a useful tool for CBIoT. }The authors of~\cite{Chall-Jour022} \textcolor{black}{have proposed a security evaluation scheme that is based on a software defined network. The network works in collaboration with twenty-three different indicators to check different elements of a clouds security. The authors break down each feature being tested in the scheme in order to display importance. Testing of the scheme shows the usefulness and the necessity of the scheme in the evaluation of CBIoT.}

	
	Moreover, the authors in~\cite{Chall-Conf022}, developed a protocol \textcolor{black}{called IoT-HiTrust}. This can be used for large mobile-cloud IoT systems. Even during disconnecting, it provides accuracy convergence and catches data securely without moving out of place, i.e., stationary.
	
	\paragraph{\textcolor{black}{Anomaly Detection}}\mbox{}
	
	The authors of~\cite{Chall-Jour023} discussed why some companies and organizations are against cloudIoT. The authors argue that cloudIoT have privacy and fewer security risks for handling huge data generated from IoT devices that are \textcolor{black}{stored} in a cloud storage with the help of SDN that enables in evaluating security level via anomaly detection.\textcolor{black}{ The authors use the ISO/IEC 25010 standard quality model to help in design their scheme.}
	
	\paragraph{\textcolor{black}{Malware Detection}}\mbox{}
	
	Malware detection strategies used to maintain the privacy of smart objects in IoT networks are \textcolor{black}{ well sought after}. So, a malware detection infrastructure is discussed in~\cite{Chall-Jour014} that consists of an Intrusion Detection System (IDS) along with cloud and fog computing. Next, a signaling game is introduced to reveal the interactions between the smart object and the corresponding fog node. To attenuate privacy leakage of smart objects, optimal strategies are also enforced that improve malware detection probability by computing the perfect Bayesian equilibrium of the game theoretically. Almost all the data present in the cloud is under the control of the Data Owner (DO), but they are out of the physical control of \textcolor{black}{the} DO, and the data in the cloud can also lead to data leakage. In this scenario, the deletion of data poses a big challenge. 
	
	\paragraph{\textcolor{black}{Intrusion Detection}}\mbox{}
	
	\textcolor{black}{As more smart devices are inserted into the realm of IoT there are more opportunities for attackers to take advantage of new entry points into a network. Intrusion detection can act as a tool to help us determine when an unauthorized user gains access to a device.} In~\cite{Chall-Jour034} the authors proposed a blockchain-aided searchable attribute-based encryption (BC-SABE) scheme with efficient revocation and decryption, for which the traditional centralized server is replaced by a decentralized blockchain system in charge of threshold parameter generation, key management, and user revocation. Also,~\cite{Chall-Conf025} proposed Edge-powered IoT Search, a new security architecture representing a collaborative and intelligent system in SDN-based cloud IoT networks, in which three layers of IDS nodes are introduced with effective node communication.

	\paragraph{\textcolor{black}{Risk Assessment}}\mbox{}
	
	In~\cite{Chall-Jour024}, a model, called SAIoT, is used for solving problems of integrated cloud and IoT with focus on risk assessment \textcolor{black}{issues}. The model consists of QoS modeling and composition. \textcolor{black}{The proposed scheme uses conditional value-at-risk to evaluate the risk assessment. A branch and check system is put in place to send inspection tasks to the system. These are both able to be done with low latency as many of the systems this is implemented in are time sensitive. }The results shown by the authors suggest that $30.64\%$ is an average improvement in QoS value.
	
	\paragraph{\textcolor{black}{Incident Response}}\mbox{}
	
	\textcolor{black}{As more IoT devices come out there is a need to monitor for incidents. The characterization of each incident can decrease the response time and make for more efficient adjustments in IoT networks. }The suggested method by authors in ~\cite{Chall-Conf011} has all of the needed security properties, especially incident response issue which are validated using the widely used BAN logic. \textcolor{black}{ The scheme uses a graph based model in order to capture knowledge about incidents for better reporting in the future.}
	
	\section{\textcolor{black}{Discussions}}\label{Disc}
	
	\subsection{\textcolor{black}{Summary of Security Challenges}}\label{Summ-Chall}
	
\begin{figure} 
		\centering
		\includegraphics[width=0.99\linewidth,keepaspectratio]{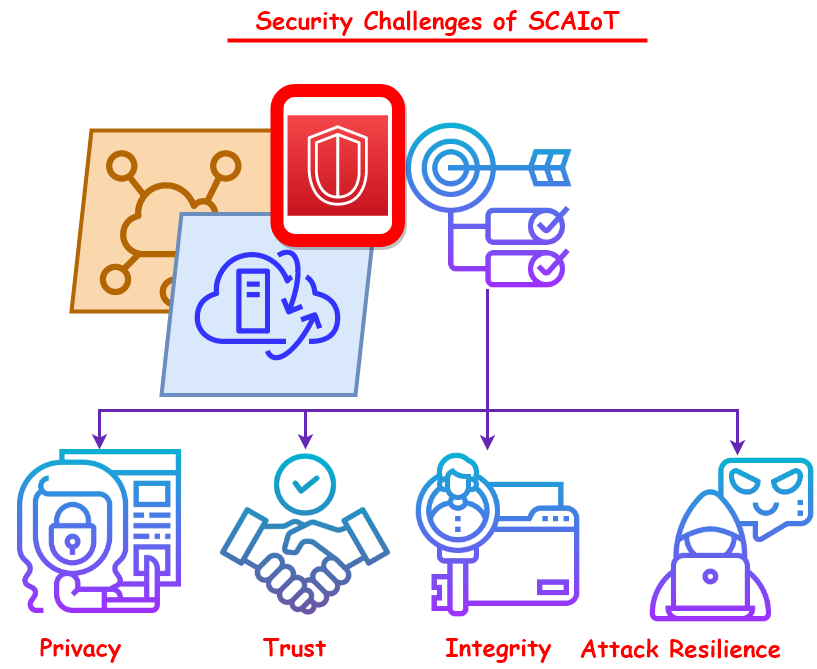}
		\caption{\textcolor{black}{Security Challenges of CAIoT}}
		\label{FigSecChal}
	\end{figure}

	\textcolor{black}{As suggested by the literature, privacy, trust, integrity, and attack resilience are the most critical challenges in the design of secure CAIoTs. These challenges are illustrated in Figure~\ref{FigSecChal}.}
	
	\subsection{\textcolor{black}{Summary of Security Controls}}\label{Summ-Cont}
	
\begin{figure} 
		\centering
		\includegraphics[width=\linewidth,keepaspectratio]{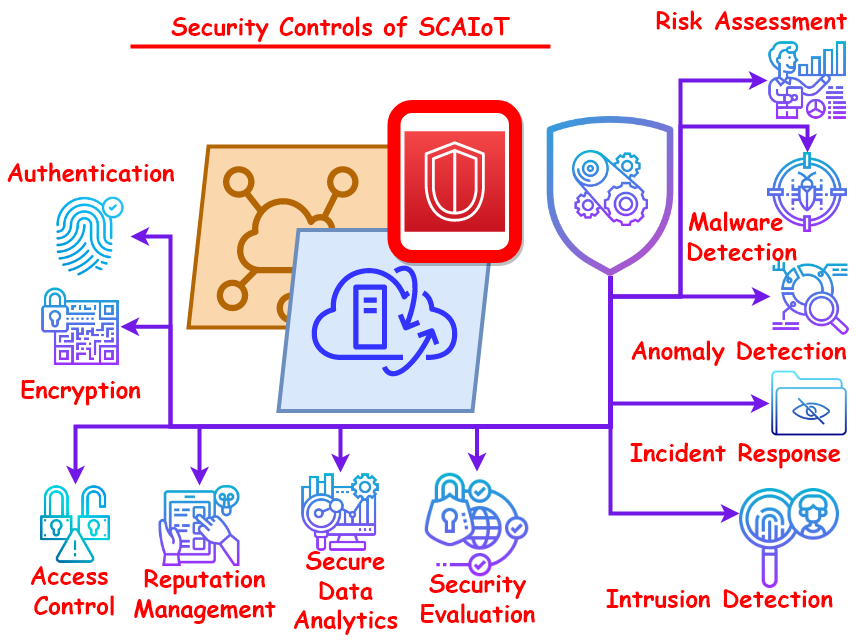}
		\caption{\textcolor{black}{Security Controls in CAIoT}}
		\label{FigSecMech}
	\end{figure}

	\textcolor{black}{Our review  identifies numerous security controls considered by designers and researchers in the field of  SCAIoT. Among these controls, one may refer to authentication, encryption, access control, reputation management, secure data analytics, security evaluation, intrusion detection \cite{rc1}, incident response, anomaly detection \cite{rc2}, malware detection and risk assessment. These security controls are depicted in Figure~\ref{FigSecMech}.}

	
	\subsection{Layered Architecture}\label{Lay-Arch}
	
\begin{figure} 
		\centering
		\includegraphics[width=0.99\linewidth,keepaspectratio]{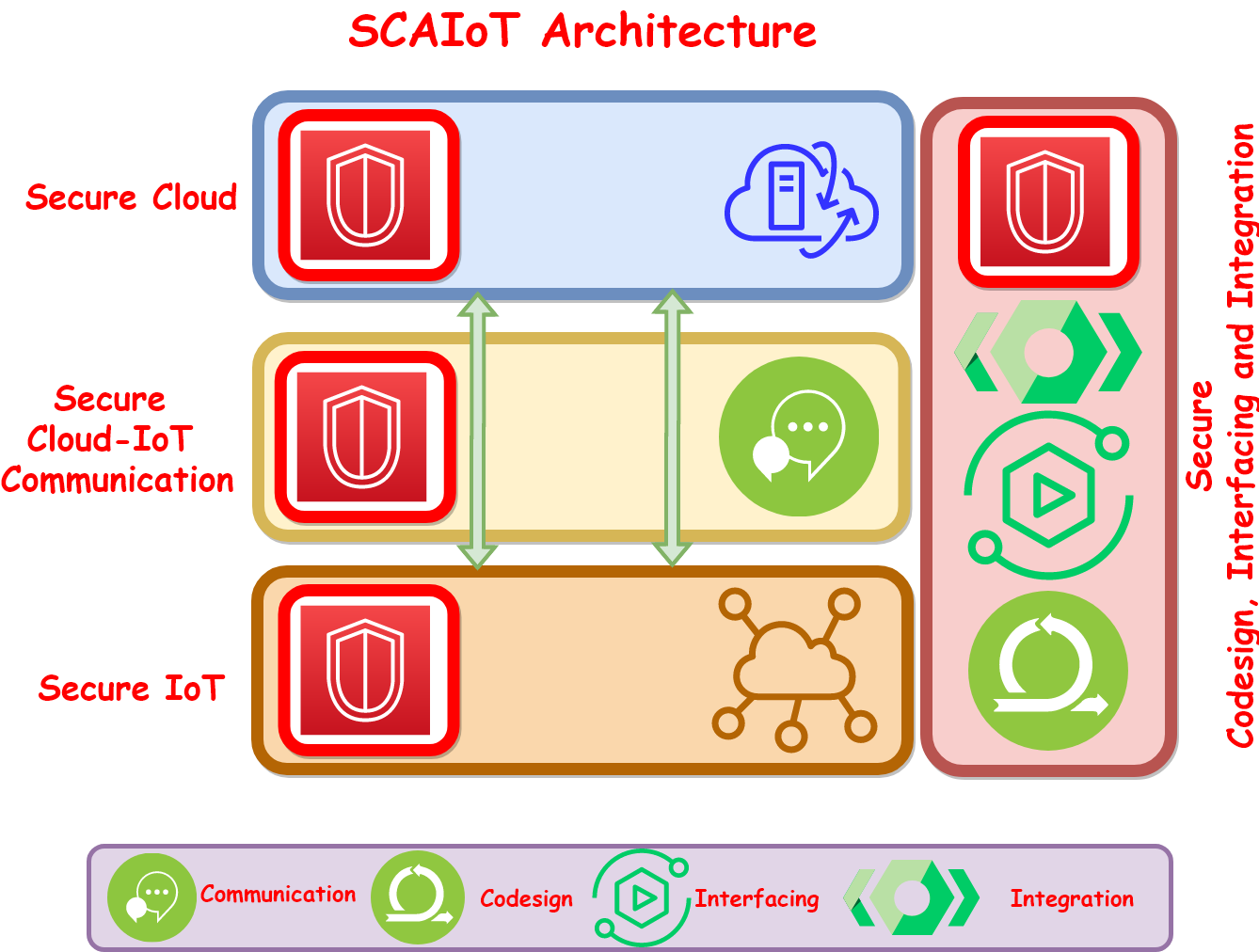}
		\caption{\textcolor{black}{The Layered Architecture of SCAIoT}}
		\label{fig:Architecure}
	\end{figure}

	\textcolor{black}{The architecture shown in Figure \ref{fig:Architecure} reflects all the approaches studied in \textcolor{black}{ Section \ref{AppChall}}.}
	
	\textcolor{black}{In Figure \ref{fig:Architecure}, the first layer is suggested by research works proposing IoT on top of secure cloud. The second layer reflects methods proposed for secure IoT-Cloud communications. Approaches based on deploying secure IoT on top of Cloud are seen in the third layer. There is also a super-layer connected to all \textcolor{black}{of the mentioned layers.} This super-layer represents approaches based on integrated SCAIoT solutions.}
	
	\section{Future Roadmap: The Promise of AI}\label{Fut}
	
	We anticipate that research on secure cloud-assisted IoT will move towards quantum-inspired AI-supported secure cloud-assisted IoT. The reason behind this anticipation is the existence of trends towards the application of AI in IoT, secure IoT, cloud, secure cloud, cloud-assisted IoT, and secure cloud-assisted IoT as well as the trend towards quantum-inspired AI, which are discussed in subsections \ref{sub1}, \ref{sub2}, \ref{sub3}, \ref{sub4}, \ref{sub5}, \ref{sub6} and \ref{sub7}, respectively.

	\subsection{AI in Cloud}\label{sub1}
	
	Several studies have been conducted in the area of AI in the cloud \cite{c2}. Tuli, et al.~\cite{Future-Jour001} proposed an Asynchronous-Advantage-Actor-Critic (A3C)-based real-time scheduler for stochastic Edge-Cloud systems that allows concurrent decentralized learning across several agents. The Residual Recurrent Neural Network (R2N2) architecture is used to capture a large number of host and task parameters and temporal trends to give efficient scheduling decisions. Next, a research work carried out by authors \cite{Future-Jour001-1} describes that a concept for the intelligent controller is suggested that connects the IoT with cloud computing and web services. Wireless sensor nodes for monitoring the indoor environment and HVAC inlet air, as well as a wireless base station for controlling HVAC actuators, have been created. 
	
	In~\cite{Future-Jour002}, the authors presented a collaborative Big.Little branch architecture to enable efficient FL for artificial intelligence IoT (AIoT) applications. Inspired by the architecture of BranchyNet, which has multiple prediction branches, their approach deploys deep neural network (DNN) models across both cloud and AIoT devices. Zhang, et al. is studying related to the ML models on cloud and trying to combine IaSS or FaSS with low cost \cite{Future-Jour003}. They propose MArk (Model Ark), a general-purpose inference serving system, to tackle the dual challenge of Service-Level Objectives (SLO) compliance and cost-effectiveness. MArk employs three design choices tailored to inference workload. Eventually, according to authors \cite{Future-Jour004}, for real-time monitoring and preventive maintenance of structural systems, a cloud-based DT framework for SHM (cDTSHM) was developed. The developed approach allows for two-way mapping between a physical structure and its digital counterpart and interaction between structure, machine, and human, paving the way for a real-time intelligent monitoring system.

	\subsection{AI in Secure Cloud}\label{sub2}
	
	The authors of~\cite{Future-Jour005} analyzed an Edge-Cloud Collaboration (ECC) scenario that includes numerous user devices with energy harvesting components, one edge server, and one cloud server\textcolor{black}{. These components} are concerned about time latency, energy consumption, and the privacy level of user devices during job offloading. The authors specifically describe the tradeoff between offloading cost and privacy level as a joint optimization issue, which \textcolor{black}{they} then model as a Markov Decision Process (MDP). \textcolor{black}{To further reduce time delay and increase security the scheme takes advantage of a deep Q-network. When tested against the widely used reinforcement learning, the proposed scheme has a reduced energy and time cost, while remaining secure.} 
	
	In \cite{Future-Jour006} the authors created SFAP, a secure federated learning technique with multiple keys that prevents DI-level poisoning assaults for medical diagnosis. Specifically, SFAP delivers privacy-preserving random forest-based federated learning through the use of secure cross computing, which protects the confidentiality of DI-related information. In another research, we can \textcolor{black}{mention TopoMAD}, a stochastic seq2seq model that can robustly model spatial and temporal dependence in contaminated data\textcolor{black}{, and} is presented by the researchers\cite{Future-Jour007}. We use system topological information to classify data from different components and use sliding windows over measures acquired constantly to capture the temporal dependence.

	\subsection{AI in IoT}\label{sub3}
	
	\textcolor{black}{AI has made many advancements in recent years with respect to cloud computing. One such example is \cite{Future-Jour008} where} the \textcolor{black}{authors} propose a unique spatial-temporal Chebyshev graph neural network model (ST-ChebNet) for traffic flow prediction to incorporate spatial-temporal data and assure reliable traffic flow prediction. \textcolor{black}{This} work proposes a novel ST-ChebNet for traffic flow prediction to capture the spatial-temporal features, which can ensure accurate traffic flow prediction. \textcolor{black}{They first} add a fully connected layer to fuse the features of traffic data into a new feature to generate a matrix, and then the long short-term memory (LSTM) model is adopted to learn traffic state changes for capturing the temporal dependencies.
	
	Researchers\cite{Future-Jour009} proposed a blockchain-based data placement protocol and theoretically model a decision optimization problem that takes into account cloud, multi-cloud, and decentralized storage technologies\textcolor{black}{. This is done} in order to select the appropriate medium to store large-scale IoT data while ensuring data integrity, traceability, auditability, and decision verifiability. In \cite{Future-Jour010}, the authors discussed CausalBG as a causal recurrent neural network (CausalRNN) deployed on an IoT platform with smartphones and CGM for precise and efficient individual blood glucose concentration prediction, according to the scientists. Another work, in \cite{Future-Jour011}, studies the security and \textcolor{black}{computes} offloading challenges in a multi-user MECCO system with blockchain at the same time. In fact, it simultaneously investigates the security and computation offloading problems in a multi-user MECCO system with blockchain. Liang, et al. apply the transfer learning approach for compressed sensing in 6G-IoT \cite{Future-Jour012}. In fact, a convolution-based transfer learning CS (CTCS) model is proposed in this research work to rebuild the compressed signal using transfer learning.

	\subsection{AI in Secure IoT}\label{sub4}
	
	In some research, AI has been adapted for the purpose of providing secure IoT \cite{c3, c4}. The authors of \cite{Future-Jour013} showed their efficient and well-developed approach for classifying risky network connections when they are no longer blacklisted or whitelisted. The solution was created for a variety of IoT cyber-physical and networking devices that were linked to the Internet. In this work, they used \textcolor{black}{a} Recurrent Neural Network (RNN) model for IoT and networking malware threat detection. Similarly, in \cite{Future-Jour014}, the authors introduce an improved deep-neural-network-based relay selection (DNS) strategy for monitoring and improving end-to-end throughput in wireless-powered cognitive Internet-of-Things (IoT) networks. In another work, \cite{Future-Jour015}, \textcolor{black}{the authors have proposed} a dynamic analysis for IoT malware detection (DAIMD) to prevent IoT device damage by detecting \textcolor{black}{ well-known and unseen and variants IoT malware} that has grown intelligently. Moreover, the authors in \cite{Future-Jour016} summarised the study and proposed a cross-architecture IoT malware threat hunting methodology based on advanced ensemble learning as an MTHAEL model.

	As another work focusing on privacy preservation, we can mention \cite{Future-Jour017}, \textcolor{black}{ where the authors have proposed} an asynchronous grouped federated learning framework (PAG-FL) for IoT, which allows numerous devices and the server to train models jointly and efficiently while maintaining privacy. The PAG-FL framework is made up of an adaptive Rényi Differential Privacy (ARB) protocol and an asynchronous weight-based grouped update (AWGU) algorithm. Also, the authors \cite{Future-Jour018} suggest DRLTrack, a framework for target tracking in Edge-IoT using collaborative deep reinforcement learning (C-DRL) to achieve high quality of tracking (QoT) and resource-efficient performance.

	\subsection{AI in Cloud-Assisted IoT}\label{sub5}
	
	Article \cite{Future-Conf001} examined the integration of cloud and edge computing for IoT data analytic and offers a deep learning-based strategy for data reduction on the edge with cloud machine learning. To decrease data dimensions, the \textcolor{black}{auto-encoder} is positioned on the edge. In \cite{Future-Jour019}, NFV, MEC, and cloud computing are combined in an NFV-enabled CECIIoT architecture. We offer an online multi-objective SFC deployment model in this architecture to balance the resource consumption and quality of various IIoT applications by describing the different service requirements and particular network environment. The authors of~\cite{Future-Conf002}, employ adaptive cloud IoT devices called Intelligent Tetris Switch (ITS) in conjunction with Big data analytic and deep learning techniques to provide smart home personalization services.
	
	\textcolor{black}{Edge computing, as described by the \textcolor{black}{authors}} \cite{Future-Conf003}, is regarded as the ideal ally for a wide range of applications for which conventional Cloud Computing is insufficient. Combining the edge method with IoT sensors and the cloud would provide consumers more freedom and options. The paper's themes provide a general open-source platform for intelligent IoT applications that is built on a shared backbone architecture that is made up of three layers: IoT objects, edge devices, and cloud infrastructure. 
	
	In \cite{Future-Conf004}, the authors researched evaluating the application of machine learning regression approaches to forecast the connection quality of communications performed by IoT nodes. Based on the node location, the suggested approach can estimate the connection quality of the most common cloud communication protocols \cite{c8,c9}, such as cellular, Wi-Fi, SigFox, and LoRaWAN. Furthrmore, we can consider JointRec in \cite{Future-Jour020}. JointRec is a deep learning-based collaborative cloud video recommendation platform, according to the author's proposal[188]. JointRec incorporates the JointCloud architecture with mobile IoT, allowing for federated training across remote cloud servers. In \cite{Future-Jour021}, the authors proposed an adaptive dropout deep computation model (ADDCM) with crowd sourcing to the cloud for industrial IoT big data feature \textcolor{black}{learning.} The article in \cite{Future-Conf001} examines the integration of cloud and edge computing for IoT data analytics and offers a deep learning-based strategy for data reduction on edge with cloud machine learning. The \textcolor{black}{auto-encoder} is positioned on edge to decrease data dimensions. 
	
	In \cite{Future-Jour019}, NFV, MEC, and cloud computing are combined in an NFV-enabled CECIIoT architecture. \textcolor{black}{The scheme offers} an online multi-objective SFC deployment model in this architecture to balance the resource consumption and quality of various IIoT applications by describing the different service requirements and particular network environments. The authors, in \cite{Future-Conf002}, employ adaptive cloud IoT devices called Intelligent Tetris Switch (ITS) in conjunction with Big data analytic and deep learning techniques to provide smart home personalization services.
	
	According to the authors of \cite{Future-Conf003}, edge computing, as described in \textcolor{black}{\cite{Future-Jour008}}, can be considered as the ideal ally for a wide range of applications for which conventional Cloud Computing is insufficient. Combining the edge method with IoT sensors and the cloud would provide consumers with more freedom and options. The paper's themes provide a general open-source platform for intelligent IoT applications built on a shared backbone architecture consisting of three layers: IoT objects, edge devices, and cloud infrastructure. Next, in \cite{Future-Conf004}, \textcolor{black}{the scheme} provides research done to evaluate the application of machine learning regression approaches to \textcolor{black}{forecast} the connection quality of communications performed by IoT nodes. Based on the node location, the suggested approach can estimate the connection quality of the most common cloud communication protocols, such as cellular, Wi-Fi, SigFox, and LoRaWAN. 
	
	Furthermore, in JointRec~\cite{Future-Jour020} a deep learning-based collaborative cloud video recommendation platform is proposed. JointRec incorporates the JointCloud architecture with mobile IoT, allowing for federated training across remote cloud servers. \textcolor{black}{The authors of} \cite{Future-Jour021}, \textcolor{black}{propose} an adaptive dropout deep computation model (ADDCM) with crowdsourcing to the cloud for industrial IoT big data feature \textcolor{black}{learning.}
	
	Based on research work in \cite{Future-Jour022}, \textcolor{black}{Cloud Video Surveillance (CVS)} is a hot topic that is widely discussed everywhere \textcolor{black}{showing the need} for real-time analysis in smart applications. There is a lack of \textcolor{black}{object} detection performance due to the complex surveilling environment. Thus in this paper, the authors have proposed multi-target object detection for real-time surveillance in IoT systems. This system is based on a deep neural network called A-YONet that is created by combining YOLO and MTCNN to build an end-edge-cloud surveillance system to realize lightweight training and feature learning with limited computing \textcolor{black}{re}sources with higher performance. Likewise, a research reported in \cite{Future-Conf005}, \textcolor{black}{has examined a dynamic} allocation strategy in the edge-cloud network over the long term with unpredictable workloads. In such a system, they present JORP, a JOint Routing and Placement problem for IoT services that dynamically distributes resources based on workload demand in order to decrease operating expenses in the long run. 
	
	In \cite{Future-Jour023}, the authors present a JOint Routing and Placement problem for IoT service chain (JORP) that can dynamically scale in/out the amount of VNF instances in this article. \textcolor{black}{Then, based on branch-and-bound(BnB), the authors use a learning approach for solving JORP efficiently}. In another work, the authors \cite{Future-Jour024} proposed a Micro-Service-based Deployment \textcolor{black}{Problem (MSDP) that} is based on the heterogeneous and dynamic characteristics of the edge-cloud hybrid environment, such as heterogeneity of edge server capacities, dynamic geographical information of IoT devices, and changing device preference for applications and complex application structures. Then, we offer a multiple buffer deep deterministic policy gradient (MB DDPG) to \textcolor{black}{ provide} more ideal service deployment options. Based on \cite{Future-Jour025}, the authors customized a federated learning framework in a cloud-edge architecture for intelligent IoT applications. To deal with heterogeneity concerns in IoT settings, we examine developing customized federated learning approaches capable of mitigating the negative impacts produced by heterogeneities in many aspects.

	With considering Energy-Efficient works in this area, the authors in \cite{Future-Jour026} propose a novel system energy consumption model that accounts for the runtime, switching, and computation energy consumption of all participating servers (from both the cloud and the edge) and IoT devices. In \cite{Future-Jour026}, to overcome the concerns above, a blockchain-based collective Q-learning (CQL) strategy is utilized \cite{Future-Jour027}, in which lightweight IoT nodes are used to train elements of training tiers, and then blockchain is used to exchange learning outcomes in a valid and essential \textcolor{black}{way} for monitoring. We view the cognitive development in the IoT node as a body of artwork rather than a pointless riddle. Additionally, in this article \cite{Future-Conf006}, the author emphasizes the need to expand beyond the domains of typical edge computing (e.g., confined to user-smartphones) and investigate ways to embed intelligence into ultra-edge IoT sensors. 
	
	In \cite{Future-Conf007}, the author employs a variety of characteristics to efficiently detect and forecast floods by utilizing cutting-edge technical equipment and concurrently transmitting through the internet, allowing us to save lives and make the world a better environment. This system makes use of several IoT devices such as a rain sensor, a water flow sensor, and a water level sensor, all of which are controlled by an Arduino microcontroller that is designed to calculate the likelihood of a flood occurring. In \cite{Future-Conf007-1}, \textcolor{black}{the authors} propose a real-time Air Quality Monitoring System (AQMS) architecture that integrates the Internet of Things (IoT) and cloud computing. AQMS operates independently and autonomously thanks to a solar panel and battery pack, making it self-powered and sustainable. Next, according to the authors in \cite{Future-Conf008}, their study presented centered on creating a comprehensive monitoring and autonomous alerting system for the elderly. By combining distinct warnings for different scenarios, the priority-based alerting system provided caregivers with a time-efficient attention method.

	\subsection{AI in Secure Cloud-Assisted IoT}\label{sub6}
	AI can address the security features of cloud-assisted IoT. As described by the \textcolor{black}{author of} \cite{Future-Conf009}, edge computing is regarded as ideal for a wide range of applications for which conventional Cloud Computing is insufficient. Combining the edge method with IoT sensors and the Cloud would provide consumers with more freedom and options. The paper's themes provide a general open-source platform for intelligent IoT applications built on a shared backbone architecture consisting of three layers: IoT objects, edge devices, and cloud infrastructure. In the article \cite{Future-Conf010}, the author talks about how introducing Cloud, IoT, ML, and AI into the system in the medical industry can increase customer satisfaction. The author has proposed a novel model that focuses on a smart hospital information management system that runs using hybrid Cloud, IoT, ML, and AI. Integration of these into their system helps the industry to customize based on customer requirements. This introduction will prove beneficial to both the medical industry and its customers. The use of unique IDs for patients and physicians would make the entire procedure much more efficient and simpler. Multi-specialty and super-specialty hospitals may create a smart hospital information management system by combining these components. The \textcolor{black}{authors introduce} a novel approach to ML services that can decrease privacy concerns while still making the services more secure. The fundamental concept is that the client delivers the partially processed feature data received from the early stage of the NN to the server while the server continues to run the rest of the NN. The privacy issue can be mitigated since it is difficult to reverse-engineer the original data from the feature data. On the other hand, the author discusses attempting to develop a privacy metric to evaluate the partially processed feature data. 
	
	Also, the major goal of this work \cite{Future-Conf012} is to offer a test technique for detecting patterns in a given dataset that do not reflect normal behavior in order to find defects, malfunctions, or the impacts of poor maintenance. This technique compares two designs for an anomaly detection system. The authors offer a real-world use case to illustrate viability. The findings of this test indicate that when Cloud computing power is increased, the full-cloud design outperforms the edge-cloud architecture and that applications may be readily assessed. 
	
	Another research work, proposed solution by the authors \textcolor{black}{of} \cite{Future-Jour028} \textcolor{black}{and} consists of an intellectual polynomial-time heuristic \textcolor{black}{scheme} that exploits the level of trust in ML models by selecting and switching between a subset of ML models from a superset of models to maximize trustworthiness while respecting the given reconfiguration budget and reducing cloud communication overhead. They also used two case studies to assess the performance of their suggested heuristic- Industrial IoT (IIoT) services and smart cities services. Similarly, the authors \textcolor{black}{of \cite{Future-Conf013} investigate} the case of data flow between the Cloud and end-user dew devices embedded into a connected \textcolor{black}{car.} This article aims to offer an IoT threat analysis and apply a deep learning technique to fight cyber anomalies, followed by a validation study of the metrics. An improvised version of Stacked Autoencoder was employed to enhance the accuracy of identifying attacks, with the loss over the training data serving as a threshold. Compared to prior models, their technique yields a better result, with 90 percent accuracy.
	
	When considering phishing detection, the authors \cite{Future-Conf014} devised a two-step process that enabled partition-based feature selection using a hybrid classifier technique and achieved optimal accuracy with a unique secure key generation mechanism, using adaptive entropy approaches. The hybrid classifier technique excelled with 97.86 percent accuracy and reduced computing complexity by 87 percent. Moreover, the authors \cite{Future-Jour029} presented a new security method called LEDEM for MDA launched by wireless IoT on an IoT server. LEDEM has a semi-supervised ML model for attack detection as well as two distinct mitigation techniques for fIoT and mIoT. When compared to the solutions, they achieved a $21\%$ gain. They claim that their security method can prevent customers from experiencing a denial of service even when the IoT server is attacked via wireless IoT. To solve cyber-attack issues, this study \cite{Future-Conf015} provides an edge-cloud deep IDS model in Lambda architecture for IoT security. When compared to standard ML methods, the introduced method minimizes training time, improves performance, and boosts the accuracy of true positive detected attacks. It may also identify suspicious behaviors in real-time and classify them by evaluating past data in a batch procedure. Likewise, this paper \cite{Future-Jour030} offers a deep blockchain framework (DBF) for IoT networks that protect from intrusion, detection, and confidentiality in blockchain with smart contracts. The suggested framework has the potential to be utilized as a decision support system to help users, and cloud providers securely migrate their data in a timely and reliable way.
	
	\begin{figure}[ht]
		\centering
		\includegraphics[width=0.99\linewidth,keepaspectratio]{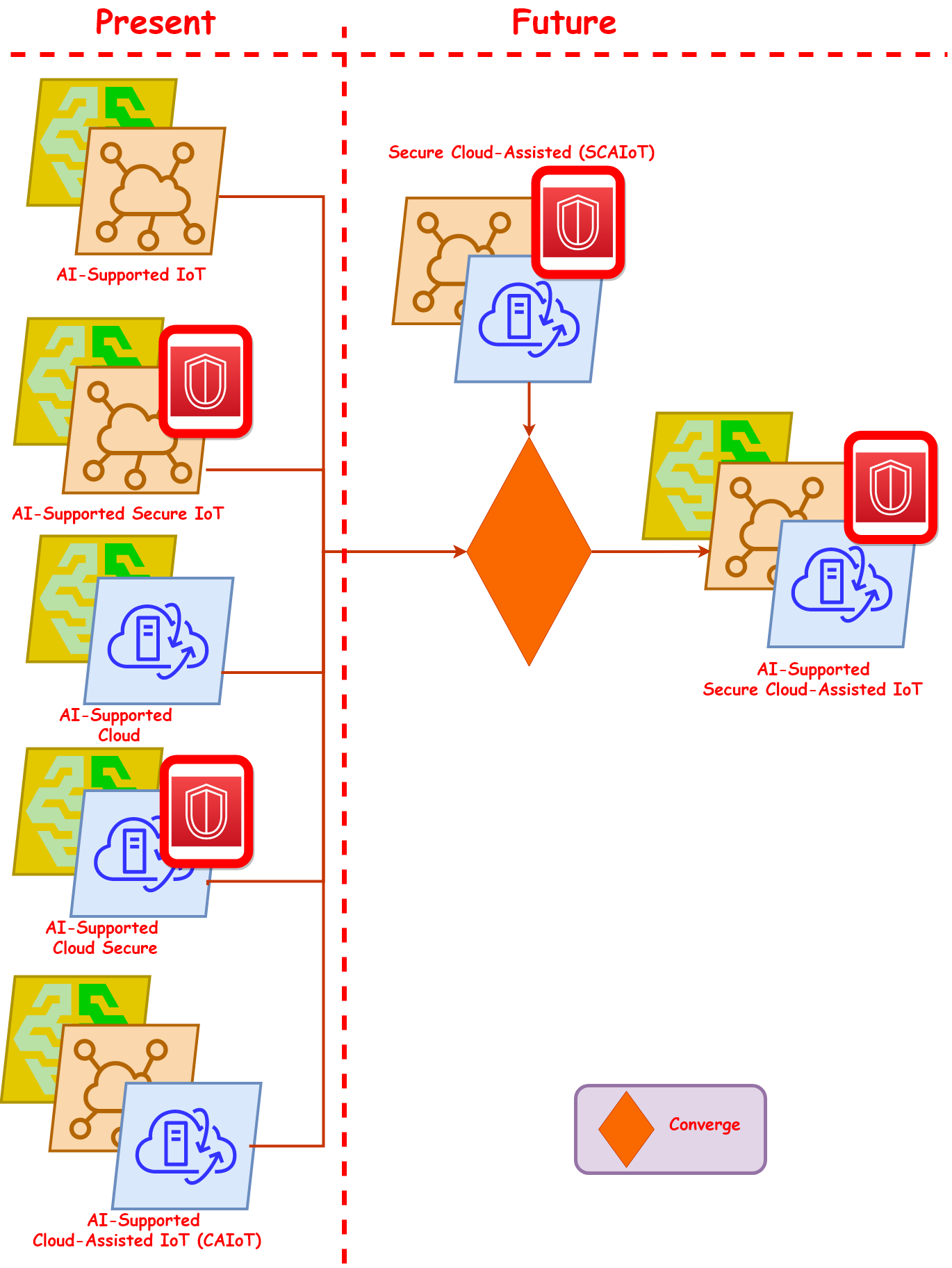}
		\caption{\textcolor{black}{The Future of SCAIoT}}
		\label{fig:Quantum Inspired AI-Supported Secure_Cloud-Assisted IoT}
	\end{figure}
	
	\subsection{Quantum-Inspired AI}\label{sub7}
	
	Researchers these days are interested in the topic of quantum AI, which is a hot topic. According to the authors \cite{Quant-AI-Jour001}, a quantum-inspired reinforcement learning (QiRL) approach to the trajectory planning problem aims to optimize the ESUTR performance for the UAV flying from the start point to the destination. Q. Wei et al.\cite{Quant-AI-Jour002} proposed a novel training paradigm inspired by quantum computation is proposed for deep reinforcement learning (DRL) with experience replay. In contrast to the traditional experience replay mechanism in DRL, the proposed DRL with quantum-inspired experience replay (DRL-QER) adaptively chooses experiences from the replay buffer according to the complexity and the replayed times of each experience (also called transition), to achieve a balance between exploration and exploitation. Based on the author's study in \cite{Quant-AI-Jour003}, quantum-inspired multi-directional association models with one-shot learning and self-convergent iterative learning, namely QMAM and IQMAM, are given in this paper. 
	
	In another work, \cite{Quant-AI-Jour004}, a novel quantum-inspired reinforcement learning (QiRL) method is suggested for autonomous robotic navigation control. The QiRL algorithm employs a probabilistic action selection policy and a novel reinforcement strategy inspired by the collapse phenomena in quantum measurement and amplitude amplification in quantum computation, respectively. To address two-class classification issues, a novel learning model dubbed Quantum-inspired Fuzzy Based Neural Network (Q-FNN) was developed by the author \cite{Quant-AI-Jour005}. The suggested model constructs the NN architecture by inserting neurons \textcolor{black}{into} the hidden layer, and learning is conducted using the idea of Fuzzy c-Means (FCM) clustering, with the fuzziness parameter (m) evolving utilizing the quantum computing concept. The quantum computing idea gives a wide search space for a selection of m, which aids in the discovery of appropriate weights as well as the optimization of network design.
	
	
	Figure \ref{fig:Quantum Inspired AI-Supported Secure_Cloud-Assisted IoT} summarizes the involvement of AI and Cloud Computing concepts with IoT. According to this figure, to make secure cloud computing and secure IoT, we need security which consists of various encryption and decryption. Further, with the development of Artificial Intelligence and combining it with the secure Cloud and secure IoT, we develop AI supported Secure CAIoT. When Neural Networks is interlinked with Artificial Intelligence, we create Quantum Inspired AI Supported Secure CAIoT.

	\section{Conclusion}\label{Conc}
	
	\textcolor{black}{This review covered some sides and aspects of the dichotomy of cloud and IoT. The dichotomy gives raise to IoT-Based Cloud (IoTBC) and Cloud-Assisted IoT (CAIoT). This paper focused on the security of CAIoT. This research identified different approaches towards the design of secure CAIoT (SCAIoT) along with the related security challenges and controls. Our reviews led to the development of a layered architecture for SCAIoT as well as a future roadmap with a focus on the role of AI. Our work in this paper can be continued via studying IoTBC or studying the unstudied aspects of CAIoT.} 

	\bibliographystyle{IEEEtran}
	\bibliography{References}
	
\end{document}